\documentclass[preprint,11pt]{elsarticle}

\usepackage[top=2.5cm,bottom=2.5cm,left=2.6cm,right=2.6cm,a4paper]{geometry}
\usepackage{graphicx}% Include figure files
\usepackage{dcolumn}% Align table columns on decimal point
\usepackage{bm}% bold math
\usepackage{epstopdf}
\usepackage{mathrsfs}
\usepackage{amsmath,amssymb,amsfonts}
\usepackage{latexsym}
\usepackage{color}
\usepackage{slashed}
\usepackage{nicefrac}
\usepackage[colorlinks=true,linktocpage=true,linkcolor=blue,citecolor=blue]{hyperref}
\usepackage[utf8]{inputenc}
\usepackage[normalem]{ulem}
\biboptions{numbers,sort&compress}

\def\bs{\boldsymbol} 
\def\del{\partial}
\def\bdel{\bs\partial}
\def\sst{\scriptscriptstyle}
\newcommand{\eqn}[1]{Eq.~\eqref{#1}}
\newcommand{\fign}[1]{Fig.~\ref{#1}}
\long\def\comment#1{ }

\newcommand{\onehalf}{{\nicefrac{1}{2}}}
\newcommand{\threehalfs}{{\nicefrac{3}{2}}}

\newcommand{\onethird}{{\nicefrac{1}{3}}}

\def\0{{\boldsymbol 0}}
\def\q{{\bm q}}
\def\l{{\boldsymbol l}}
\def\k{{\boldsymbol k}}

\def\n{{\boldsymbol n}}
\def\x{{\boldsymbol x}}
\def\y{{\boldsymbol y}}
\def\p{{\boldsymbol p}}

\def\r{{\boldsymbol r}}
\def\z{{\boldsymbol z}}
\def\u{{\boldsymbol u}}
\def\v{{\boldsymbol v}}

\def\Q{{\boldsymbol Q}}

\def\Mmat{\mathsf{M}}

\def\Rmat{\mathsf{R}}

\def\llangle{\left \langle}
\def\rrangle{\right \rangle}
\def\beps{{\boldsymbol \epsilon}}

\def\sM{\text{med}}

\def\tmat{ \text{\bf t}}
\def\tmat{ \textbf{t}}
\def\Tmat{ \text{\bf T}}
\def\ti{ t}
\def\tf{t'}
\def\tend{{t_\text{\tiny L}}}
\def\tform{{t_\text{f}}}
\def\tdecoh{t_\text{d}}
\def\tquant{t_\text{quant}}
\def\tres{t_\text{res}}

\def\tbr{t_\text{br}}

\def\Gc{{\cal G}}

\def\Mc{{\cal M}}

\def\Wc{{\cal W}}
\def\Dc{{\cal D}}
\def\Kc{{\cal K}}

\def\ProbOne{P_{\scriptscriptstyle 1}}
\def\ProbOneM{\tilde P_{\sst 1}}

\def\ProbTwo{ P_{\scriptscriptstyle 2} }
\def\ProbSing{ P_\text{sing}}
\def\ProbTwoM{\tilde P_{\scriptscriptstyle 2}}
\def\ProbSingM{\tilde P_\text{sing}}
\def\pT{p_{\sst T}}

\newcommand{\beq}{\begin{eqnarray}}
\newcommand{\eeq}{\end{eqnarray}}
\newcommand{\be}{\begin{eqnarray*}}
\newcommand{\ee}{\end{eqnarray*}}
\newcommand{\bal}{\begin{align}}
\newcommand{\eal}{\end{align}}
\newcommand{\rmd}{{\rm d}}
\newcommand{\dd}{{\rm d}}
\newcommand{\rme}{{\rm e}}

\newcommand{\rmtr}{{\rm tr}}
\newcommand{\rmTr}{{\rm Tr}}
\def\rmR{{\rm Re}}

\def\abar{{\rm \bar\alpha}}

\newcommand{\nn}{\nonumber\\ }

\newcommand{\labe}{\label}

\begin{document}

\begin{frontmatter}

\title{Radiative energy loss of neighboring subjets}

\author[int]{Yacine Mehtar-Tani}
\ead{ymehtar@uw.edu}
\author[cern]{Konrad Tywoniuk}
\ead{konrad.tywoniuk@cern.ch}

\address[int]{%
Institute for Nuclear Theory, University of Washington, Box 351550,\\ Seattle, WA 98195-1550, USA
}
\address[cern]{%
Theoretical Physics Department, CERN, 1211 Geneva 23, Switzerland
}
\date{\today}% It is always \today, today,

\begin{abstract}
We compute the in-medium energy loss probability distribution of two neighboring subjets at leading order, in the large-$N_c$ approximation. Our result exhibits a gradual onset of color decoherence of the system and accounts for two expected limiting cases. When the angular separation is smaller than the characteristic angle for medium-induced radiation, the two-pronged substructure lose energy coherently as a single color charge, namely that of the parent parton. At large angular separation the two subjets lose energy independently. Our result is a first step towards quantifying effects of energy loss as a result of the fluctuation of the multi-parton jet substructure and therefore goes beyond the standard approach to jet quenching based on single parton energy loss.  We briefly discuss applications to jet observables in heavy-ion collisions. 
\end{abstract}

\begin{keyword}
perturbative QCD \sep jet physics \sep jet quenching
\end{keyword}

\end{frontmatter}

\begin{flushright}
INT-PUB-17-022, CERN-TH-2017-134
\end{flushright}

%%%%%%%%%%%%%%%%%%%%%%%%%%%%%
\section{Introduction}
\label{sec:introduction}
%%%%%%%%%%%%%%%%%%%%%%%%%%%%%

First observed at RHIC \cite{Adcox:2001jp,Adler:2002xw}, then at LHC \cite{Aamodt:2010jd,CMS:2012aa,Aad:2015wga}, the large suppression of high-$\pT$ particle spectra in nucleus-nucleus collisions, referred to as ``jet quenching", is commonly regarded as a signature of the formation of the quark-gluon plasma (QGP). This striking phenomenon is attributed to medium-induced radiative energy loss of high momentum partons as they propagate through a hot and spatially extended medium \cite{Bjorken:1982tu,Gyulassy:1990ye,Wang:1991xy}. 
Measurements of fully reconstructed jets allowed the investigation of new  jet quenching observables covering not only measurements of the suppression of inclusive jet spectra \cite{Aad:2010bu,Chatrchyan:2012nia,Aad:2014bxa,Abelev:2013kqa} and photon-jet correlations \cite{Chatrchyan:2012gt}, but also including information about the redistribution of jet energy \cite{Khachatryan:2015lha} within and outside the jet cone \cite{Chatrchyan:2012gw,Chatrchyan:2013kwa,Aad:2014wha}. 
In consequence, jet observables, and jet substructure measurements in particular, have shed new light on the mechanism of jet-medium interactions and the details of energy loss while posing new challenges to the theoretical description of in-medium fragmentation at the same time.
 
In-medium jet quenching have so far been treated at the level of single parton energy loss whose radiation pattern is affected by multiple scattering. This dynamical process gives rise to the Landau-Pomeranchuk-Migdal (LPM) intereference \cite{Baier:1994bd,Baier:1996sk,Baier:1996kr,Baier:1998yf,Zakharov:1996fv,Zakharov:1997uu,Wiedemann:2000za}, that causes the suppression of the radiation spectrum at large frequencies. The dominant soft emissions take place at time-scales parametrically smaller than the medium size and can therefore be treated as quasi-instantaneous \cite{Blaizot:2012fh,Apolinario:2014csa}. This permits a probabilistic treatment of multiple emissions \cite{Blaizot:2013vha}. For soft gluon emission, the resulting cascade is governed by turbulence \cite{Blaizot:2013hx} which efficiently transports energy to large angles \cite{Blaizot:2014ula,Blaizot:2014rla,Kurkela:2014tla,Iancu:2015uja}. 
 
The probability of emitting a total energy $\epsilon$ off the leading parton passing through a medium, $\ProbOne(\epsilon)$, was first discussed in \cite{Baier:2001yt}. In the absence of any angular constraints, this probability distribution is governed by the energy scale $ \omega_s \sim \alpha_s^2 \,\hat q\, L^2$, in a medium of size $L$ and characterized by the jet quenching parameter $\hat q$, which is a diffusion coefficient in transverse momentum space. It follows that the related fluctuations of energy loss are also of the same order \cite{Escobedo:2016jbm,Escobedo:2016vba}. The resultant single-parton spectrum is shifted towards decreasing energies as compared to the primordial one leading to an overall suppression (see also \cite{Arleo:2002kh,Salgado:2003gb,Baier:2006fr}). See also \cite{Arnold:2015qya,Arnold:2016kek,Arnold:2016mth,Arnold:2016jnq} for related work on medium-induced multi-gluon emissions.

Extending the calculations of radiative energy loss from single partons to jets proved to be a difficult problem. 
In point of fact, there is a large probability for nascent partons to branch due to the infrared and mass singularities of QCD splittings, and many jet properties, such as the effect of the cone size, jet substructure and jet mass, appear only on the level of at least one splitting.
Moreover, it is difficult to argue that all splittings take place outside of the medium, typically extending over several fm's, which would justify a treatment where only the parent parton suffers energy loss. This implies that jets propagate through the medium as multi-parton quantum states whose energy loss pattern is expected to differ from that of independent partons. Moreover, it is well know that color coherence plays an important role and leads to angular ordering of subsequent emissions \cite{Mueller:1981ex,Ermolaev:1981cm}. These probabilistic features arise due to the active role of interferences. 
The treatment of jet fragmentation vertices inside the medium remains therefore an open question. Nevertheless, several Monte-Carlo prescriptions that are based on heuristic arguments exist in the literature, see e.g. \cite{Young:2011ug,Zapp:2012ak}.

\begin{figure}[t!]
\begin{center}
\includegraphics[width=0.8\textwidth]{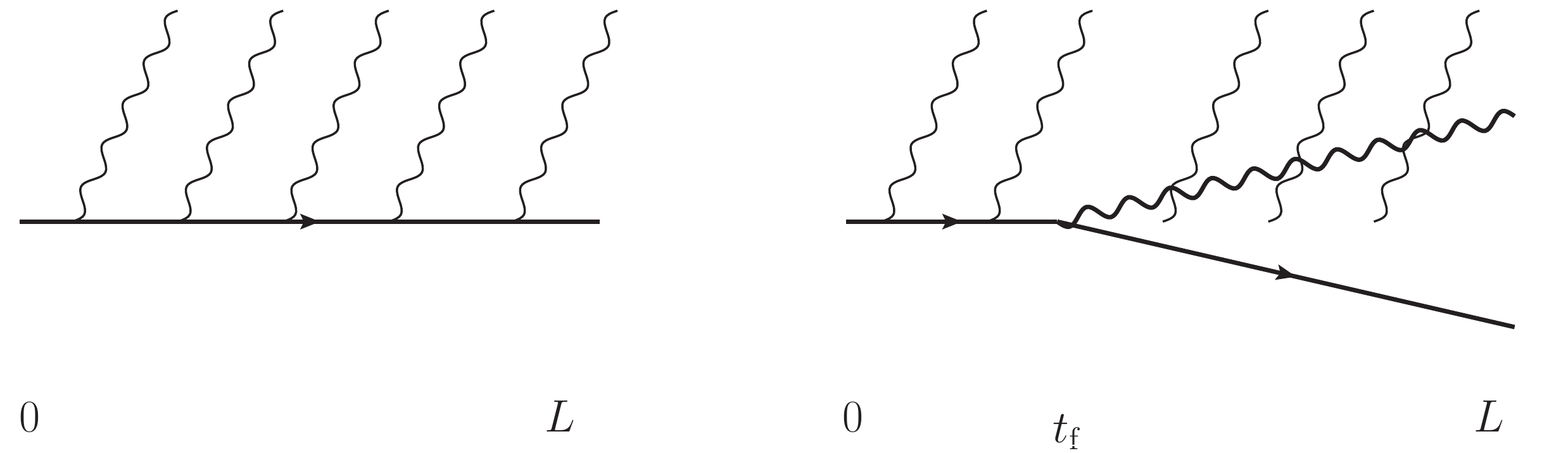}
\caption{Radiative energy loss for a single particle (left) and a two-pronged color object (right).}
\label{fig1}
\end{center}
\end{figure}

The purpose of this paper is to address this problem from first principles. The key point of our formalism is  the resummation of multiple soft primary radiation off a color dipole with proper treatment of interferences. The processes under consideration are depicted in \fign{fig1}. In contrast to single-particle energy loss (left), having two participants (right) begs the question of how energy is collectively removed from the system. 
The main part of this work will consider this problem in detail and arrive at a description of how a system of two partons, formed early in the medium, are affected by radiative energy-loss. In consequence, this extends the scope of \cite{Baier:2001yt} (which focuses on the problem of energy-loss of a single charge) and provides a novel tool for dealing with jet observables. We call the resulting distribution the {\sl two-prong} quenching weight, which will be analyzed in the large-$N_c$ limit. 

Having in mind a phenomenological observable, our calculation deals with the spectrum of two-pronged substructures inside a jet, both prongs being significantly more energetic than any medium scale. This corresponds to relatively symmetric splitting configurations with small formation times (hard emissions). 
For the moment, we assume the offspring to be sufficiently energetic so that their momenta and energies are not further modified by medium rescattering, thus, treating the medium-induced energy loss as a small correction compared to the final subjet energies. 
Although the kinematics of the hard splitting is unaltered, within our approximations, their corresponding yield, however, is affected 
as a result of the convolution of the energy loss probability with the steeply falling spectrum of the hard process.
A direct measurement of this mechanism could be facilitated by a wide range of jet substructure techniques, for reviews see \cite{Sapeta:2015gee,Salam:2009jx,Altheimer:2012mn}, and was already considered as a measure of medium effects in \cite{Mehtar-Tani:2016aco}.

%%%%%%%%%%%%%%%%%%%%%%%%%%%%%
%%%%%%%%%%%%%%%%%%%%%%%%%%%%%

Our findings substantiate an intuitive picture for the two-pronged energy loss. Confirming earlier expectations \cite{CasalderreySolana:2011rz,MehtarTani:2011gf,MehtarTani:2012cy}, the dynamical effects due to
the finite resolution power of medium fluctuations play an important role.
Initially after formation, which is assumed to be instantaneous, the pair remains close in transverse distance and is not resolved by the medium. This implies that energy is taken coherently away from the pair and this process is governed by the total color charge, i.e. the medium sees the pair {\it as the parent parton}. There are two effects that break the coherence of the pair. The first is active close to the origin of the splitting and is common to radiative processes in vacuum and medium \cite{CasalderreySolana:2011rz,MehtarTani:2012cy}. 
Recall that a dipole can only give rise to radiation that is sensitive to its moment, i.e. with transverse wavelength $\lambda_\perp \sim k_\perp^{-1} < x_{\perp}$, where for a time-like dipole the size grows linearly with time $x_\perp \sim \theta_{12} t$. However, at a given time $t$, only certain quanta have developed quantum-mechanically.
Substituting $t$ for the typical formation time of a splitting $\tform \sim \omega/k_\perp^2$, transforms this conditions into the well-known angular constraint $\theta < \theta_{12}$, which reflects the fact that large angle radiation is strongly suppressed due to color coherence. 
The second effect, which dominates at large times, is related to the loss of color coherence of the dipole constituents due multiple interactions with the medium that cause their rapid color randomization. It takes place when the medium resolution scale, $(\hat q  t)^{-1/2}$ which is related to the in-medium transverse momentum broadening,  is smaller than the dipole size $x_{12}(t) $.  That is, $\theta_{12}  t > (\hat q  t)^{-1/2}$, that yields the characteristic color decoherence time 
\beq 
\tdecoh \sim (\hat q \theta^2_{12})^{-1/3}.
\eeq
Throughout, we shall ignore vacuum radiation and focus on medium-induced radiation. At its formation in the medium, the gluonic fluctuation had accumulated a transverse momentum $k_\perp^2\sim \hat q \tform $.\footnote{For hard medium interactions, or thin media, the typical resolution scale is related to the transverse momentum exchange with a single constituent. In this case however, the antenna is only partially resolved, for more details see \cite{MehtarTani:2011gf}.} Recalling that $\tform \sim \omega /k_\perp^2$, we can readily solve for $\tform$ and obtain the characteristic medium-induced formation time $\tform \sim \sqrt{\omega/\hat q } $ and the corresponding radiation angle $\theta_\text{f}\sim k_\perp/\omega\sim (\hat q /\omega^3)^{1/4}$. 
This soft medium-induced radiation can be triggered anywhere along the medium with $\tform  \ll t < L $, in contrast to vacuum radiation for which $t \sim \tform$. In the regime of interest, $\tdecoh \gg \tform$,  the decoherence time sets the transition between coherent radiation for which $ t \ll \tdecoh$ and independent radiation when $ t \gg \tdecoh$. 
In the situation where  $\tform \gg \tdecoh$ on the other hand, quantum decoherence dictates when interferences are suppressed but this occurs for medium-induced emission at angles that are smaller than $\theta_{12}$ as can be seen  after re-expressing the color decoherence time  as $ \tdecoh \sim \tform \,(\theta_{12}/\theta_\text{f})^{2/3}$.
Here an important remark is in order. We will be interested in radiative energy loss of a dipole as a proxy for a jet that is defined by the sum of all particles within a given cone, which we assume here to be $\theta_{12}$.  This implies that only radiation at angles larger than $\theta_{12}$ will contribute to the energy loss of the system. Hence, quantum decoherence that is only active, as we showed above, at angles smaller than  $\theta_{12}$ can be neglected and only color decoherence can lead to independent radiation at large angles. In other words, angular ordered radiation does not contribute to energy loss.

Let us now return to the justification of the instantaneous generation of the dipole. We have assumed that the dipole formation is vacuum-like and hence occurs at times $t_\text{f(dipole)} \ll t_\text{d(dipole)} $.
This condition determines the kinematical regime where we can treat the vacuum splitting separately from subsequent medium modifications (in our case, energy loss). In addition, we have that $t_\text{f(dipole)} \ll L$ since we are interested in splittings inside the medium. Furthermore, in order to justify the application of the ``quenching'' approximation, namely the fact that we will only consider primary emissions off the hard particles, amounts to demanding that the energies of the pair constituents is much larger that $\omega_c =\hat q L^2$ the characteristic frequency of medium-induced radiation. At sufficiently high energies, this condition is violated only for very asymmetric splittings where we have to include the possibility of generating the pair via medium-induced processes. Related arguments have recently been raised in the context of the single-inclusive jet spectrum and the corresponding phase space was found to be large \cite{Mehtar-Tani:2017web}.

Furthermore, requiring that $\theta_{12} \gg \theta_c$ which corresponds to $t_\text{d}  \ll L$ (the regime of color decoherence), implies that $\omega_c \equiv \hat q L^2  \gg \sqrt{\hat q L} / \theta_{12} $. This condition together with the assumption $E \gg \omega_c$ yields a constraint on the transverse momentum of the pair $E \theta_{12}\gg \sqrt{\hat q L}$, to be larger that the transverse momentum broadening in the medium which can therefore be neglected. In the opposite case, $\theta_{12} \ll \theta_c$, the pair is not resolved by the medium and hence, energy is lost coherently.  As a result, we do not expect momentum broadening to play a significant role in the limit $E \gg \omega_c$.

A corollary to our calculation is to consider for the first time radiative corrections to the 
jet quenching parameter that affects the dipole in the color coherence regime.
It was recently realized that both the processes of transverse momentum broadening and gluon emission will receive logarithmically enhanced radiative corrections due to long-lived fluctuations in the medium \cite{Liou:2013qya,Blaizot:2014bha,Wu:2014nca,Iancu:2014kga}. It was shown that those universal corrections can be reabsorbed in a renormalization of the jet quenching parameter. We are able to demonstrate that this also holds for the related process of color decoherence.

Let us also note that while our calculations are applied to the case of a dense medium, where multiple scattering dominates, it can also be extended to dilute media. In the latter case, interference effects between energetic partons were already analyzed in \cite{MehtarTani:2011gf}, where it was found that the level of decoherence is proportional to the opacity of the medium. Therefore, in order to achieve completely independent energy loss, multiple scatterings are necessary.

We structure our paper as follows. In order to set the stage, we give a in-depth discussion on how energy is lost by propagating color charges in the medium in Sec.~\ref{sec:PartonEnergyLoss}. We recapitulate the energy loss distribution for a single color charge and go on to present our main results for the two-prong energy loss distribution, in particular \eqn{eq:main-result} for the color singlet antenna and \eqn{eq:main-result-colored} for the generalization to arbitrary color representation.
The details of single parton energy loss, presented in Sec.~\ref{sec:QuenchingWeights}, allows us to introduce our formalism. The full derivation of the two-pronged energy loss distribution is presented in Sec.~\ref{sec:derivation}. In particular, the radiative corrections appearing for the process of color decoherence are derived in Sec.~\ref{sec:radiative-corrections}. Finally, we summarize and present an outlook in Sec.~\ref{sec:conclusions}. Our
Feynman rules are listed in 
\ref{sec:FeynmanRules}, and the remaining Appendices contain further details of the  calculations. A final remark, we work in light-cone (LC) coordinates, defined as $x^+ = (x^0 + x^3)/2$ and $x^- = x^0 - x^3$, and will abbreviate the LC $+$-momentum as $p^+ \equiv E$ and LC-time as $x^+ \equiv t$.

%%%%%%%%%%%%%%%%%%%%%%%%%%%%%
\section{Summary of the results}
\label{sec:PartonEnergyLoss}
%%%%%%%%%%%%%%%%%%%%%%%%%%%%%

In order to help the reader walk through the steps of our derivation, we explicitly restate here our working assumptions at the outset:
\begin{enumerate}[(i)]
\item The energetic antenna forms instantaneously at $t=0$, for a discussion see \ref{sec:HardVertex}. 
\item The radiated gluons are very soft such that the total energy loss is much smaller than the energies of either of the two prongs. 
\item After the formation of the antenna, only medium-induced radiation is considered, see \ref{sec:n-point-fct} for details. 
\item We neglect overlapping medium-induced emissions owing to the fact that each of their individual formation times are much smaller than the medium length.
\item Finally, we  focus on the leading-$N_c$ contribution.
\end{enumerate}

%%%%%%%%%%%%%%%%%%%%%%%%%%%%%
\subsection{One-parton energy loss probability}
\label{sec:one-prong-solution}
%%%%%%%%%%%%%%%%%%%%%%%%%%%%%

Let us first review the problem of radiative energy loss of a single, high energy parton passing through a static, dense QCD medium of length $L$. For simplicity, let us for the moment assume it being a quark. In this setup, the medium scales are a small correction, hence $\omega \ll E$, where the energy $E$ refers to any projectile parton and $\omega$ to the medium-induced gluons. In fact, as we will shortly convince ourselves, medium interactions provide a maximum energy scale $\sim \omega_c$, which allows us to justify the high-energy limit.

In this limit the emissions spectrum does not depend explicitly on the parent energy,
and the cross-section factorizes as follows:
\beq
\frac{\rmd \sigma }{\rmd \omega \rmd E} \simeq   \frac{\rmd I}{\rmd \omega} \, \frac{\rmd \sigma_\text{vac}}{\rmd E'}  \,,
\eeq
where $\sigma_\text{vac}$ is the quark cross-section and $ \rmd I / \rmd \omega $ stands for the medium-induced bremsstrahlung spectrum that we will discuss shortly. Here, $E'=E+\omega$ is the quark energy before radiation. The gluon frequency is neglected everywhere except in $\sigma_\text{vac}$, since the steepness of the spectrum allows a small energy loss to yield a large effect on the final spectrum. Typically, $\dd \sigma_\text{vac}/\dd E \propto E^{-n}$ where $n \gg 1$, hence the relevant scale that probes energy loss is shifted to $\epsilon \sim E/n \ll E$ \cite{Baier:2001yt}.

To find the inclusive spectrum of single quarks in heavy-ion collisions, one has to integrate over the gluon frequency and add virtual corrections. 
The final-state spectrum of quarks in heavy-ion collisions is then given as a convolution of the (vacuum) production spectrum $\dd \sigma_\text{vac}/\dd E$ with the subsequent probability of losing energy,
\beq
\label{eq:one-prong-factorization}
\frac{\dd \sigma_\text{med}}{\dd E} = \int_0^{\infty}\dd \epsilon\, \ProbOne(\epsilon)\frac{\dd \sigma_\text{vac}(E+\epsilon)}{\dd E'} \,.
\eeq
where have introduced the energy loss probability distribution which, to leading order in $\alpha_s$ is given by 
\beq\labe{eq:e-loss-LO}
\ProbOne(\epsilon)\simeq \delta(\epsilon)    \left(1 - \int_0^\infty \rmd \omega\frac{\rmd I}{\rmd \omega}  \right)+  \frac{\rmd I}{\rmd \epsilon}\,,
\eeq
The complete derivation is presented in Sec.~\ref{sec:QuenchingWeights}. In \eqn{eq:e-loss-LO}, the first term corresponds to the probability for the high energy parton not to lose energy, namely, that no radiation takes place. The third term corresponds to real emissions while the second term, inside the parenthesis, corresponds to a virtual correction that ensures probability conservation.
 
The medium-induced soft spectrum that enters \eqn{eq:e-loss-LO} accounts for multiple scattering in the medium, and can be found from the rate \cite{Baier:1994bd,Baier:1996sk,Zakharov:1996fv,Zakharov:1997uu},
\beq
\label{eq:LPM-rate}
\frac{\dd I }{\dd \omega \dd t} = \abar \sqrt\frac{\hat q}{\omega^3} \,,
\eeq
where  $\bar \alpha \equiv \alpha_s C_F/\pi$. The rate of emissions off a gluon projectile is found by substituting the color factor $C_F \to C_A$.
This spectrum reflects the fact that the typical transverse momentum of the emitted gluon is given by a diffusion process, $\langle k_\perp^2 \rangle \sim \hat q t$, where $\hat q$ is the jet transport coefficient and $t$ the distance travelled in the medium. 
The gluon formation time therefore becomes $\tform \sim \sqrt{\omega / \hat q}$, which is limited by the size of the medium. Hence \eqn{eq:LPM-rate} is only valid for gluon energies $\omega \lesssim \omega_c \equiv \hat q L^2/2$. Finite-size corrections, neglected so far in this discussion, strongly suppresses the spectrum for $\epsilon \ll \omega_c$ \cite{Baier:1994bd,Baier:1996sk,Zakharov:1996fv,Zakharov:1997uu}. 

The probability in \eqn{eq:e-loss-LO} corresponds to a single emission. Higher order corrections are important as can be seen by computing the gluon multiplicity above $\omega$, 
\beq
\label{eq:bdmps-multiplicity}
N(\omega) = \int_0^L \dd t \int_\omega^\infty  \dd \omega'\, \frac{\dd I }{ \dd \omega' \dd t} = 2\bar \alpha\, \sqrt{ \frac{\hat q L^2}{\omega}}\,.
\eeq
In effect, $N(\omega) >  1$ for $\omega <  \omega_s$ where $\omega_s\sim \bar\alpha^2 \hat q L^2  \ll \omega_c$. In this regime, therefore, multiple emissions ought to be resummed. Nonetheless, soft gluon emissions can be regarded as quasi-instantaneous, $\tform\ll L$ for $\omega \ll \omega_c$, and treated as independent \cite{Blaizot:2012fh,Apolinario:2014csa}. 

As we shall see in more details in the next section, this amounts to writing the single parton energy loss probability as resummed product of independent medium-induced emissions, i.e.
\beq
\label{eq:quenching-weight-1}
\ProbOne(\epsilon) =\Delta(L) \, \sum_{n=0}^\infty \,\frac{1}{n!} \, \prod_{i=1}^n \, \int_0^L\rmd t   \int \rmd \omega_i \,  \frac{\rmd I}{\rmd \omega_i \rmd t  }  \, \delta\left(\epsilon-\sum_{i=1}^n\omega_i\right) \,,
\eeq
where $\epsilon$ corresponds to the total energy emitted and the Sudakov-like factor,
\beq
\label{eq:Sudakov}
\Delta(L)\equiv \exp\left( -\int_0^L\rmd t \int_0^\infty \rmd \omega\frac{\rmd I}{\rmd \omega \rmd t} \right) \,,
\eeq
represents the probability of not radiating between 0 and $L$. This leads to the famous expression for the radiative energy loss probability
\beq
\label{eq:one-prong-eloss}
\ProbOne(\epsilon) = \sqrt{\frac{2\omega_s}{\epsilon^3}} \, \,\rme^{-\frac{2\pi \omega_s}{\epsilon}} \,,
\eeq
where $\ProbOne(\epsilon) \equiv \ProbOne (\epsilon,L)$, 
which is only sensitive to the soft energy scale $\omega_s \equiv \bar \alpha^2 \omega_c$, so long as $\omega \ll \omega_c$ \cite{Baier:2001yt}, see also \cite{Arleo:2002kh,Salgado:2003gb,Baier:2006fr}. 
The probability distribution in \eqn{eq:one-prong-eloss} is usually referred to as the quenching weight and forms the basis of most theoretical studies of jet quenching, see e.g. \cite{Andres:2016iys,Arleo:2017ntr}. The main underlying assumption of these studies is that the initial projectile does not split inside the medium with a vacuum probability. For high-energy jets, this requirement neglects the contributions coming from hard radiation that can be formed early in the medium. Calculating how energy-loss affects such processes is the topic of the next subsection.

%%%%%%%%%%%%%%%%%%%%%%%%%%%%%
\subsection{Two-parton energy loss probability at large-$N_c$}
\label{sec:two-prong-solution}
%%%%%%%%%%%%%%%%%%%%%%%%%%%%%

We shall now address the main question of this article, namely,  the energy loss probability for a neighboring pair of hard partons that originate from the same vertex. For the moment, we disregard any modification to the formation of the pair and study their subsequent evolution. Hence, we shall assume that the splitting occurs quasi-instantaneously so that one can factor out the Born level from the rest of the process. 
Such a setup extends the concept of the quenching weights, described in the previous subsection, and is a step toward investigating the problem of energy loss of full-fledged jets. The two partons are color connected and formed in an arbitrary (singlet, triplet or octet) color representation of SU(3). 

As for single-parton quenching, we shall compute the medium modification of the two-parton system spectrum. In the collinear limit, $p_\perp \ll E$, the quark-gluon spectrum in vacuum is given by 
\beq\label{eq:vac-2prong-spect}
\frac{\dd\sigma_\text{vac}}{\dd z \dd E \,\dd \p^2} \simeq \frac{\alpha_s C_F}{2\pi} \frac{P_{gq}(z)}{\p^2} \frac{\dd \sigma_\text{vac}}{\dd E} \,,
\eeq
to leading order in perturbation theory and at leading logarithm, where we denote $p_\perp \equiv |\p|$ throughout the paper.  In \eqn{eq:vac-2prong-spect}, $P_{gq}(z)$ stands for the quark-gluon  Altarelli-Parisi splitting function and $\dd \sigma_\text{vac} \big/\dd E$ is the Born-level quark spectrum.

For a hard instantaneous splitting the  two-parton distribution in the presence of a medium  reads 
\beq
\label{eq:two-prong-spectrum}
\frac{\dd \sigma_\text{med}}{ \dd z \dd E \,\dd \p^2} = \int_0^\infty \dd \epsilon \, \ProbTwo(\epsilon) \, \frac{\dd\sigma_\text{vac}}{\dd z \, \dd E' \,\dd \p^2} \,,
\eeq
analogously to Eq.~(\ref{eq:one-prong-factorization}), where  $\ProbTwo(\epsilon)\equiv \ProbTwo^{\sst R}(\epsilon,\theta_{12},L)$ stands for the energy loss probability distribution for a two-prong structure in the color representation $R$. In this case, the opening angle is $\theta_{12} = p_\perp/[z(1-z)E]$.
This approximation holds when there is large separation between the time-scales of production of the hard pair and the time-scale of medium modifications. In particular, this involves the case $\tform \ll \tdecoh$, which contains the leading logarithmic phase space, for a further discussion see also \cite{Mehtar-Tani:2017web}. 
We leave a calculation of finite-energy corrections for the future.

The only dependence on $\epsilon$ in the spectrum appears in the parent parton spectrum, where $E'=E+\epsilon$. The splitting function $P(z)$ is unmodified because the correction goes as $\Delta P(z) = P'(z) \epsilon/E$ \cite{Mehtar-Tani:2016aco}, which is suppressed as long as the energies of the daughter partons are large enough, $zE \gg \epsilon$. However, due to the steeply falling spectrum, which we assume to be governed by a power law behavior, as above,
the ratio of medium and vacuum spectra is more sensitive to modifications
so the typical energy loss should again be compared to  $\epsilon \sim E/n \ll  E $ and, hence, cannot be neglected \cite{Baier:2001yt}. 

For reasons that will shortly become clear, the singlet antenna is the irreducible color configuration that one needs to consider in the large-$N_c$ limit in order to construct the general result for arbitrary color states. Hence, we shall focus on the decay of a boosted massive object in a singlet color state such as a virtual photon decaying into a quark-antiquark ($q\bar q$) state, neglecting for the moment the quark masses.
Finally, we consider the same set of approximations as for the single quark case in the previous subsection. In particular with regards to the treatment of radiation, we shall work in the soft gluon approximation for medium-induced gluons, corresponding to quasi-instantaneous emissions, where overlapping formation times are neglected, see the discussion in Sec.~\ref{sec:QuenchingWeights}. Then, as a final step we will then generalize this description to account for a generic total charge of the parent parton.

%%%%%%%%%%%%%%%%%%%%%%%%%%%%%
\begin{figure}[t!]
\begin{center}
\includegraphics[width=0.65\textwidth]{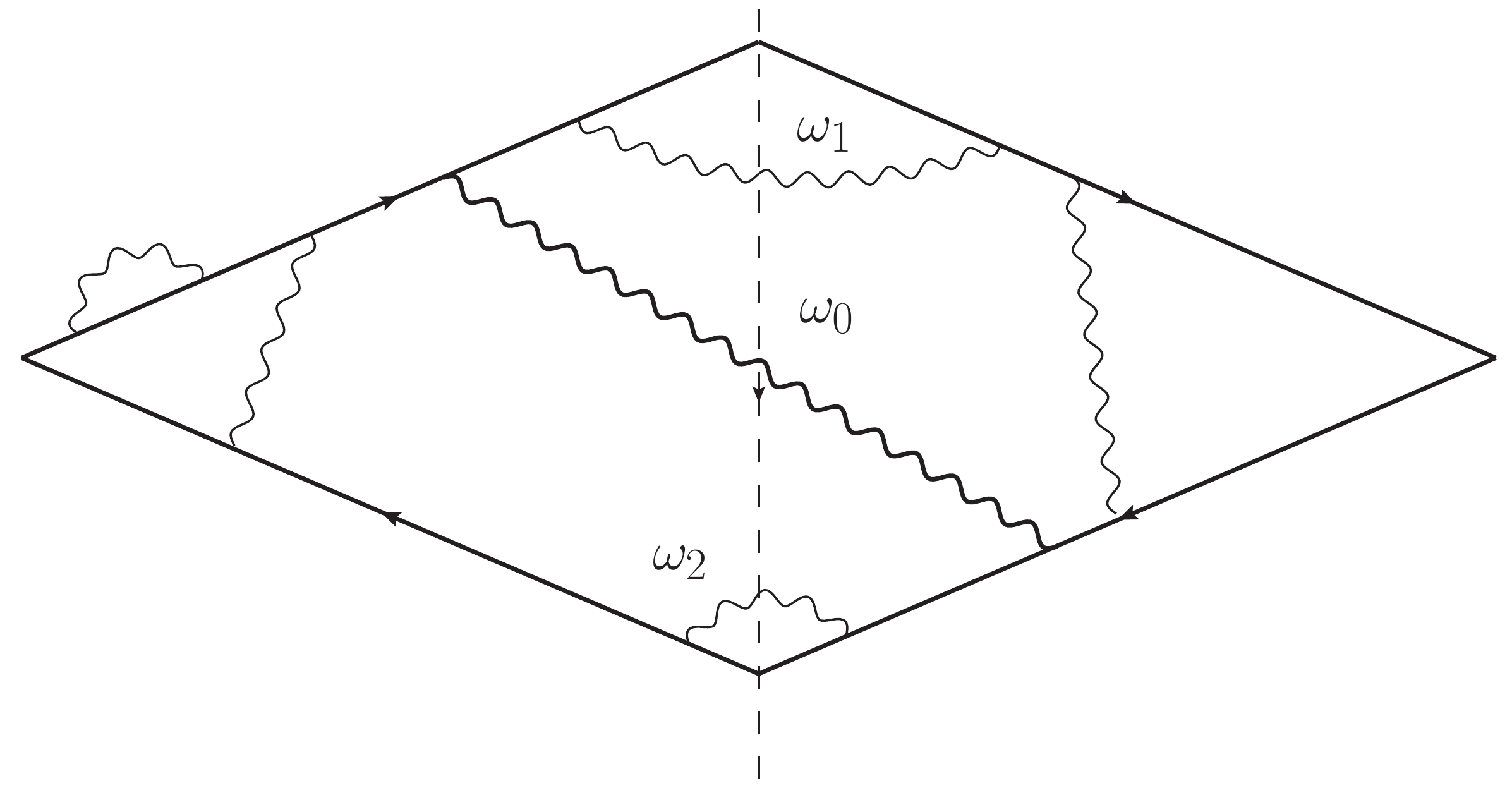}
\caption{Feynman diagram depiction of the process of decoherence and subsequent energy loss of two partons created early in the medium. In the presence of the two-prong structure, we integrate over all medium-induced emissions, depicted by wavy lines, that can occur as direct contributions (e.g., such as the gluons labeled by $\omega_1$ and $\omega_2$), interferences (e.g., the gluon labeled by $\omega_0$) and virtual contributions.}
\label{fig5}
\end{center}
\end{figure}
%%%%%%%%%%%%%%%%%%%%%%%%%%%%%

The cross section of the generic process we are interested in is depicted in the standard form in \fign{fig5}, where the amplitude (and its complex conjugate) is depicted on the left (right) side of the cut, represented by the dashed line, and we integrate over all medium-induced gluons, depicted by wavy lines.
The crucial observation that reduces the complexity of the task at hand consists in realizing that there can only be one gluon connecting the quark (antiquark) in the amplitude and the antiquark (quark) in the complex conjugate amplitude in the large-$N_c$ limit  \cite{Dominguez:2012ad}, as represented by a thick wavy line labeled by $\omega_0$ in \fign{fig5}. This emission constitutes at the same time a modification of the color structure of the antenna, see Sec.~\ref{sec:interference}, which is why we refer to it as a ``flip''. Hence, to the left of the flip in the amplitude, 
only virtual diagrams contribute which, as we will see in Sec.~\ref{sec:radiative-corrections}, will lead to the renormalization of the quenching parameter $\hat q$ inside the decoherence parameter.  To the right of the flip radiation off the quark and the antiquark factorize, resulting in independent radiative energy loss, as denoted by the thin, gluon lines labeled $\omega_1$ and $\omega_2$ in \fign{fig5}. 

%%%%%%%%%%%%%%%%%%%%%%%%%%%%%
\begin{figure}[t!]
\begin{center}
\includegraphics[width=0.9\textwidth]{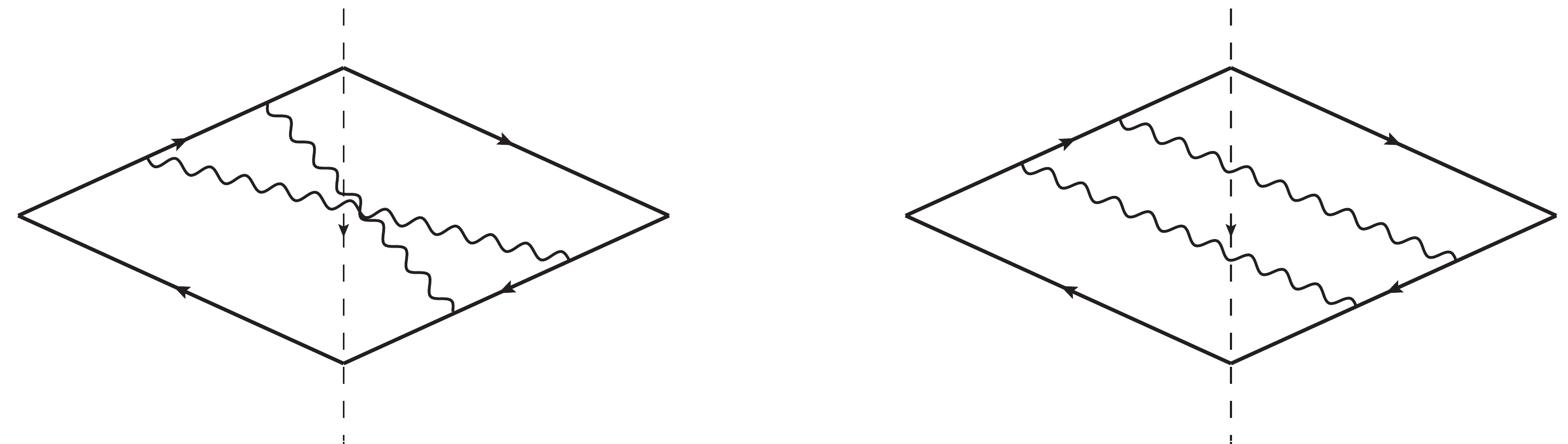}
\caption{Example diagrams of sub-leading contributions that are neglected in our treatment. The left diagram is non-planar and, thus, suppressed in the large-$N_c$ limit, and the right diagram is suppressed because of overlapping formation times. }
\label{fig7}
\end{center}
\end{figure}
%%%%%%%%%%%%%%%%%%%%%%%%%%%%%

This intuitive picture emerges within our set of approximations that make away with more complicated situations and topologies that are sub-leading. Two such interference contributions are depicted in \fign{fig7}. The diagram on the left constitutes a non-planar contribution, which is explicitly neglected in the large-$N_c$. The diagram on the right is leading in $N_c$. Nevertheless, it corresponds to two-gluon emission with overlapping formation times. This is sub-leading in the limit of soft gluon emissions, due to the limited phase space for overlapping formation times in a large medium, see e.g. \cite{Blaizot:2013vha}. 

Let us currently report the result of our calculation for the color singlet energy loss distribution, leaving a full derivation to Sec.~\ref{sec:derivation} (see also Sec.~\ref{sec:QuenchingWeights} for an introduction to the general formalism). A generalization to arbitrary color representation will follow shortly. We define a regularized splitting rate, acting on the propagating quark-antiquark system as
\beq
\label{eq:gamma-ij}
\Gamma_{ij}(\omega, t) \equiv \frac{\rmd I_{ij}}{\rmd \omega\, \dd t} -\delta(\omega)\int_0^\infty  \rmd \omega' \frac{\rmd I_{ij}}{\rmd \omega'\, \dd t} \,,
\eeq
where the index $i$ ($j$) denotes the leg that is emitting (absorbing) the medium-induced gluon (regardless whether it is in the amplitude or complex conjugate amplitude). 
In the following, the index ``1'' (``2'') will refer to the quark (antiquark). For example, $\n_1 = \p_1/E_1$ ($\n_2 = \p_2/E_2$) refers to the direction of propagation of the quark (antiquark), and so on. We find that, 
\begin{align}
\label{eq:bdmps-direct}
\frac{\dd I_{11}}{\dd \omega \dd t} = \frac{\dd I_{22}}{\dd \omega \dd t}& = \abar \sqrt{\frac{\hat q}{\omega^3}} \,,\\
\label{eq:bdmps-interference}
\frac{\dd I_{12}}{\dd \omega \dd t} = \frac{\dd I_{21}}{\dd \omega \dd t}& = - \abar \sqrt{\frac{\hat q}{\omega^3}} \,\textsl{F}\left(t/t_\text{quant}\right) \,,
\end{align}
for the direct and interference spectra, respectively. The function $\textsl{F}\left(x\right) \sim\Theta(1-x)$, for details see discussion leading to \eqn{eq:spectrum-intermediate}, incorporates the effect of quantum decoherence which suppresses the interference term for hard vacuum emissions that resolve the dipole when they are formed at times $t  > t_\text{quant} \sim (\theta_{12}^2 \omega)^{-1} \sim (\theta_\text{f}/\theta_{12})^{4/3} \,\tdecoh $. For details, see Sec.~\ref{sec:interference}. In the regime of interest, that is for medium-induced radiation at angles larger than the opening angle $\theta_\text{f} \gg \theta_{12}$, we have $\tdecoh  \ll t_\text{quant} $.\footnote{In order to quantify this statement further, we point out that radiation inside the cone occurs with probability $\mathcal{O}(\alpha_s)$ as long as $\omega_s \ll (\hat q/\theta_{12}^4)^{1/3} < \omega_c$.} Therefore, the mechanism of color decoherence is active before quantum decoherence (see the introduction for a similar discussion). One can thus neglect the latter by letting $F(x) \sim 1$. 
As a result, the interference spectra (stripped of the decoherence parameter (\ref{eq:decoh-parameter}))  are approximately equal to the direct emission spectra, 
\beq
\frac{\dd I_{11}}{\dd \omega \dd t} \simeq \frac{\dd I_{22}}{\dd \omega \dd t}\simeq -\frac{\dd I_{12}}{\dd \omega \dd t}\simeq -\frac{\dd I_{21}}{\dd \omega \dd t}\simeq \frac{\dd I}{\dd \omega \dd t}.
\eeq
Putting all the pieces together,  the two-prong energy loss probability of a color singlet dipole in the large-$N_c$ limit reads
\begin{align}
\label{eq:main-result}
\ProbSing(\epsilon) &= \int_0^\infty\rmd \epsilon_1 \int_0^\infty \rmd \epsilon_2 \,  \ProbOne(\epsilon_1, L) \, \ProbOne(\epsilon_2, L)\, \delta(\epsilon-\epsilon_1-\epsilon_2) \nn 
& - 2 \int_0^L \rmd t  \int_0^\infty\rmd \epsilon_1 \int_0^\infty \rmd \epsilon_2 \, \ProbOne(\epsilon_1, L-t) \, \ProbOne(\epsilon_2, L-t) \nn
& \times \,  \Big[ 1- \Delta_\sM(t) \Big] \int_0^\infty \rmd \omega  \, \Gamma (\omega, t) \, \delta(\epsilon-\epsilon_1-\epsilon_2-\omega) \,,
\end{align}
where $\ProbSing(\epsilon) \equiv \ProbSing(\epsilon,\theta_{12},L)$. The factor $-2$ in the last line results from the sum over the two contributions to the interference spectrum $\sum_{i \neq j}\Gamma_{ij} (\omega, t) \approx -2 \Gamma (\omega, t)$.
This is one of the main results of this paper.
Here, $\ProbOne(\epsilon,L)$ describes the independent energy loss of the antenna legs, see \eqn{eq:one-prong-eloss}, and $\Delta_\text{med}(t)$ is the so-called decoherence parameter 
that incorporates the effect of color decoherence, and
reads (for a homogeneous medium) 
\beq
\label{eq:decoh-parameter}
\Delta_\text{med}(t) = 1- \exp \left[ - \frac{1}{12} \hat q \, \theta^2_{12}\,  t^3 \right]  \,,
\eeq
see Sec.~\ref{sec:radiative-corrections} for further details. It constitutes the probability for the medium to resolve the color structure of the pair after traversing a distance $t$ in the medium and is sensitive to characteristic angle $ \sim (\hat q t^3)^{-\onehalf}$, above which color coherence is wiped out.
One can check that \eqn{eq:main-result} contains two limiting cases, corresponding to coherent and decoherent antennas. In order to illustrate this point, let us for the moment focus exclusively on the effect of color decoherence, contained in the decoherence parameter \eqn{eq:decoh-parameter}. First, we deal with the incoherent case. Letting $\Delta_\sM(L) =1$, i.e. $\theta_{12} \gg \theta_c$ where $\theta_c \sim (\hat q L^3)^{-\onehalf}$, suppresses the second term and therefore the total energy loss probability of the singlet two-prong structure is given by the product of one-prong energy loss probabilities,
\beq\labe{eq:main-result-decoh}
\ProbSing(\epsilon) \simeq \int_0^\infty\rmd \epsilon_1 \int_0^\infty \rmd \epsilon_2 \,  \ProbOne(\epsilon_1) \, \ProbOne(\epsilon_2)\, \delta(\epsilon-\epsilon_1-\epsilon_2) \,.
\eeq 
In the opposite case $\theta_{12}=0$, so that  $\Delta_\sM(t) =0$. In this limit the interferences cancel the direct contributions, $\rmd I_{12} \simeq -\, \rmd I_{11}$.\footnote{Strictly speaking, this is true for soft, medium-induced emissions that are not affected by angular ordering. } In the interference term, proportional to $\Gamma( \omega,t)$, we can shift $\epsilon_1\to \epsilon_1-\omega$ to allow the integral over $\omega$ to act on $\ProbOne(\epsilon_1-\omega, L-t)$.
Then one can use that, cf. \eqn{eq:eloss-rate-eq},
\beq
\frac{\del }{\del t } \ProbOne(\epsilon_1, L-t) = -  \int \rmd \omega\,  \Gamma (\omega,t)    \ProbOne (\epsilon_1-\omega, L-t) \,,
\eeq
and similarly with $\ProbOne(\epsilon_2, L-t)$. We reconstruct in this way the total derivative acting on the product of energy-loss probabilities, such that the interference term gives rise to 
\beq
\int_0^L  \rmd t\, \frac{\del }{\del t }  \big[ \ProbOne (\epsilon_1, L-t) \, \ProbOne(\epsilon_2, L-t)\big] = \delta(\epsilon_1) \delta(\epsilon_2)-  \ProbOne(\epsilon_1, L) \, \ProbOne(\epsilon_2, L),
\eeq
where the second term cancels exactly the first term in \eqn{eq:main-result-decoh}. Therefore, the energy loss probability of the infinitely narrow singlet antenna vanishes, that is when $\theta_{12} \ll \theta_c$, reads 
\beq
\ProbSing(\epsilon) \simeq  \delta(\epsilon) \,,
\eeq
 as expected.

The generalization to arbitrary color representation of the parent parton is straightforward in the large-$N_c$ limit, see Sec.~\ref{sec:final-answer} for details. Let us for the moment consider parton splittings via a gluon emission.\footnote{This does not directly apply to $g \to q+\bar q$ splittings which do not involve any interferences in the large-$N_c$ limit and which are anyway suppressed in the leading-logarithmic approximation.}
Hence, for a parent parton with color representation $R$, the two-pronged energy loss probability reads,
\beq
\label{eq:main-result-colored}
\ProbTwo^{\sst R}(\epsilon) = \int_0^\infty \dd \epsilon_1 \int_0^\infty \dd \epsilon_2 \,\ProbOne^{\sst R}(\epsilon_1) \ProbSing(\epsilon_2) \delta(\epsilon - \epsilon_1 - \epsilon_2) \,,
\eeq
where $\ProbSing(\epsilon)$ denotes the color-singlet two-prong quenching weight in \eqn{eq:main-result}.
Keep however in mind that the color factor in the interference spectrum entering $\ProbSing(\epsilon)$ has to be replaced by $C_F \approx N_c/2$ and the one-prong quenching weight $\ProbOne^{\sst R}(\epsilon)$ becomes sensitive to the relevant color charge through the generalized coupling $\bar \alpha_{\sst R}= \alpha_s C_R/\pi$. This is the second main result of the paper.

%%%%%%%%%%%%%%%%%%%%%%%%%%%%%
\begin{figure}[t!]
\begin{center}
\includegraphics[width=0.48\textwidth]{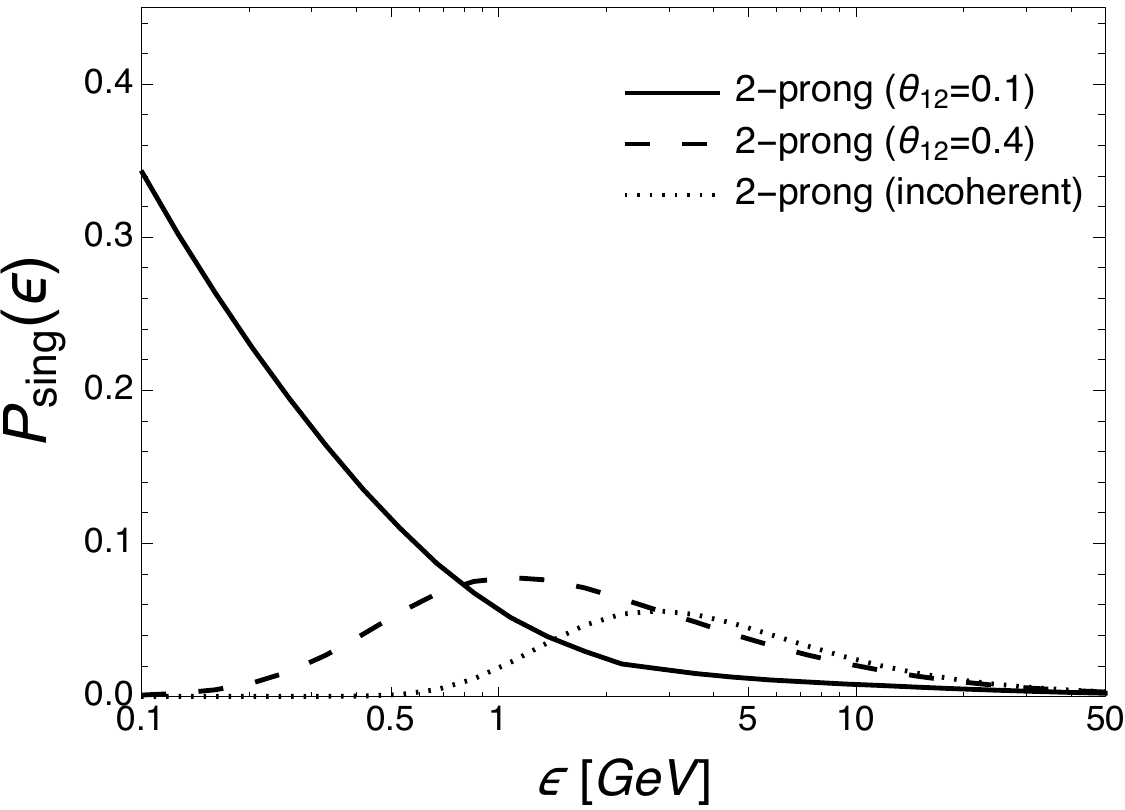}
\includegraphics[width=0.48\textwidth]{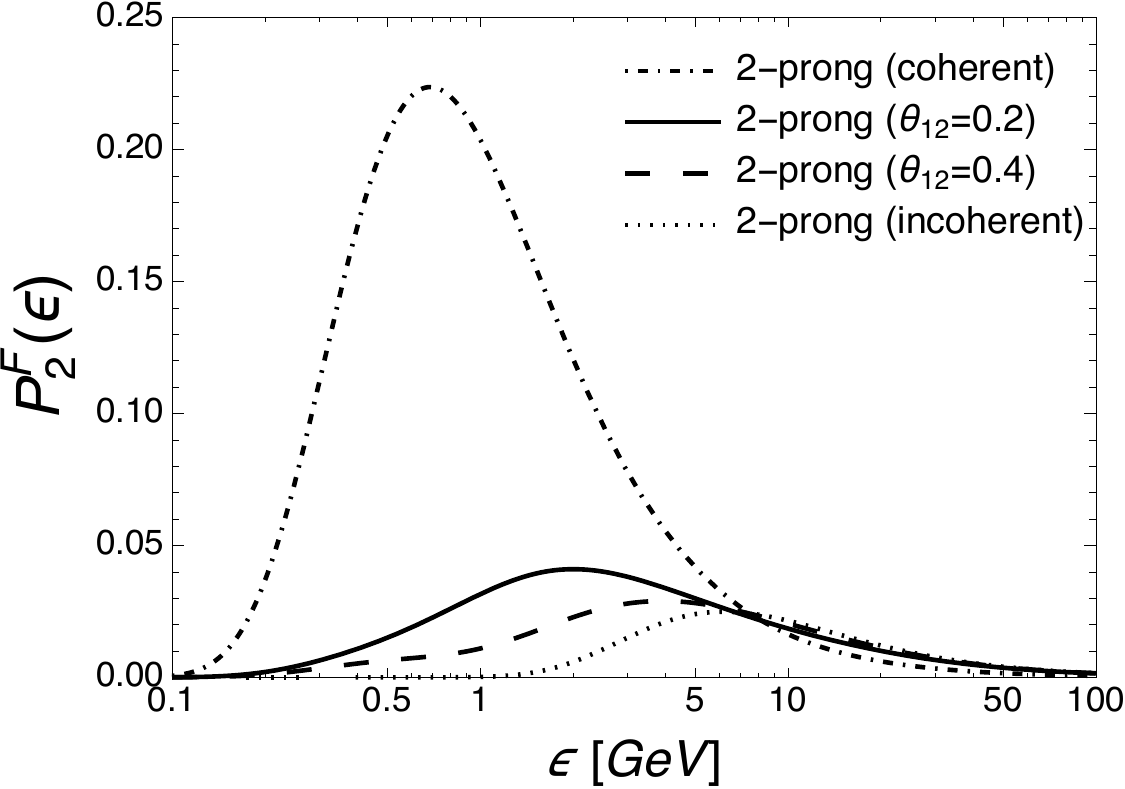} 
\caption{The two-prong energy loss distribution for a color singlet antenna (left) and for a quark splitting (right). Medium parameters were chosen to be $\hat q = 1$ GeV$^2$/fm and $L=2$ fm, and $\alpha_s=0.3$. We plot the corresponding distributions for two opening angles, $\theta_{12} = 0.2$ (solid curves) and $\theta_{12}=0.8$ (dashed curves) in addition to the completely incoherent case (dotted curves) and, for the quark, the one-prong energy loss distribution corresponding to a completely coherent splitting (dash-dotted curve).}
\label{figEloss}
\end{center}
\end{figure}
%%%%%%%%%%%%%%%%%%%%%%%%%%%%%

We plot the two-prong energy loss distribution for a color singlet antenna, evaluated according to \eqn{eq:main-result}, and for a quark splitting, evaluated according to \eqn{eq:main-result-colored}, in \fign{figEloss} for two different opening angles, see figure caption for details.
For comparison, we also evaluate a two-pronged incoherent energy loss distribution, which simply is defined by the first line in \eqn{eq:main-result}. Finally, for the quark we also plot the single-prong energy loss distributions that corresponds to a completely coherent splitting.
The color singlet distribution interpolates between a distribution peaked around small values of $\epsilon$ for small opening angles to the incoherent case at large angles. We notice immediately the sensitivity to the opening angle. At sufficiently high energy, all distributions fall off as $\epsilon^{-\threehalfs}$, due to the incoherent nature of hard emissions.

To summarize, for arbitrary color representation in the large-$N_c$ limit, the two-pronged energy loss distribution is therefore a convolution of the quenching weight of the total charge along the whole length of the medium with the two-pronged color singlet distribution that we will derive in full detail in the subsequent sections.

%%%%%%%%%%%%%%%%%%%%%%%%%%%%%
\section{Single parton energy loss}
\label{sec:QuenchingWeights}
%%%%%%%%%%%%%%%%%%%%%%%%%%%%%

In this section we shall detail the calculation of the single parton energy loss probability $\ProbOne$ discussed in Sec.~\ref{sec:one-prong-solution}. It will serve to introduce the formalism that we shall use in the next section where we derive the main result of this work, namely, the two-prong energy loss probability distribution. 

The parton, of energy $E$ and transverse momentum $\p$, is assumed to be produced in a hard process, such as an $e^+e^-$ collision. We will keep the initial angle finite, but small $ |\p|/E <1$, although one can choose a frame where it is zero, in order to introduce the most general formulation that will be used in the case of two color charges. 
In fact, as we will show below, the radiation spectrum will not depend on the angle. In this case, emissions off the recoiling quark can be ignored by choosing the light-cone gauge $A^+=0$, which ensures that soft gluons can only be radiated by a forward-moving quark (antiquark). In the present analysis, we also ignore bremsstrahlung radiation associated with the creation of the back-to-back pair and focus solely on medium-induced radiation.
 
We are interested in medium effects within the following set of approximations:
(i) we neglect the transverse momentum broadening of the quarks, and (ii) we assume soft gluon emission, $\omega \ll E$. In this limit, the amplitude (up to irrelevant phase factors) reads,
\begin{align}
\label{eq:amp-mixed-3}
\Mc^{(a,i)}_{(\lambda,s)}(p,k) &= \frac{g}{\omega} \int_0^L \rmd t \, \, \rme^{i\frac{\omega  }{2 }\n^2 t}\,\,  (\bdel_x + i \omega \n ) \cdot\beps^\ast_\lambda\, \left. \Gc_{_A}^{ab} (\k,L; \x, t ) \right|_{\x=\n t} \nn
&\times  \left[  V(L,t)    \tmat^b\,  V(t, 0)  \right]^{ij}   \Mc^{j}_{s}(p) \,,
\end{align}
where  $s$ and $\lambda$ are the spin and polarization of the quark and the radiated gluon, respectively.
For details see \ref{sec:HardVertex}. In this expression, $\Gc_{_A}$ describes the propagation of the gluon and $V(\tf,\ti) \equiv U_{_F}\big(\tf,\ti; [\x]\big)$ stands for a Wilson-line in the fundamental representation evaluated along the classical trajectory $\x$, 
given by $\x(s)=\n s$ where $\n\equiv \p/E$, see \eqn{eq:propagator-quark}. Note that the initial point of the eikonal gluon propagator $ \Gc_{_A}^{ab} (\k,L; \x, t )$ is evaluated in coordinate space while the endpoint in momentum space that is,
\beq
\Gc_{_A}^{ab} (\k,L; \x, t ) =\int \rmd^2 \y  \, \,\Gc_{_A}^{ab} (\y,L; \x, t ) \,\, \rme^{-i \y\cdot \k}\,.
\eeq
Apart from the color structure, $\Gc$ is equivalent to the Green's function of a non-relativistic particle of mass $\omega$ in 2+1 dimensions propagating through a background field, described by the Schr\"odinger equation 
\beq\label{eq:prog-schr}
\left[i\frac{\del}{\del t} + \frac{{\bs \del}^2}{2 \omega} + g \mathcal{A}(\x,t)\right] \, \Gc(\x,t; \x_0,t_0) = i \delta(\tf-\ti) \, \delta(\x-\x_0) \,,
\eeq
where $\mathcal{A} \equiv \Tmat\cdot A^-$ and $\Tmat$ is the relevant color matrix. The solution can be cast in the form of a path integral,
\beq
\label{eq:greens-function}
\Gc(\x,t; \x_0,t_0)  = \Theta\big(t -t_0 \big) \int_{\r(t_0)=\x_0}^{\r(t)=\x} \mathcal{D} \r \exp \left[i\frac{\omega}{2} \int_{t_0}^{t} \dd s \, \dot \r^2 (s)\right] \,U\left(t,t_0; [\r] \right) \,,
\eeq
where the Wilson line reads,
\beq
\label{eq:propagator-quark}
U\big( t,t_0;[ \r] \big) = \mathcal{P}_+ \, \exp \left[ ig \int_{t_0}^{t} \dd s\, \mathcal{A}\big(\r(s),s \big) \right] \,,
\eeq
and $\mathcal{P}_+$ implements path ordering.

The differential cross section is obtained by squaring the amplitude, averaging over initial and summing over the final state quantum numbers (helicity, color, flavor) and background field configurations. Since we are only interested in the inclusive energy spectrum, we also can integrate out the transverse momentum of the emitted gluons. The cross section then reads
\beq
\label{eq:amplitude-squaring}
\frac{\rmd \sigma }{\rmd \Omega_k \,\rmd \Omega_p } = \, n_f \Big\langle \frac{1}{N_c}\sum_{\lambda, s}  |\Mc^{(a,i)}_{(\lambda,s)}(p,k)|^2 \Big\rangle \,,
\eeq
where $n_f$ is the number of quark flavors and the factor $1/N_c$ corresponds to an average of initial quark colors, and the gluon measure reads,
\beq  
\rmd\Omega_k\equiv  \frac{1}{4\pi}\frac{\rmd \omega}{\omega}\frac{\rmd^2 \k }{ (2\pi)^2},
\eeq 
and similarly for the quark. Here, $\langle ...\rangle$ stands for the ensemble average over medium configurations, which is assumed to be Gaussian with the 2-point correlator given by 
\beq\label{eq:med-average}
\langle  {\cal A}^a (\q,t){\cal A}^{\ast a} (\q',t')\rangle \equiv n\, \delta^{ab}\, (2\pi)^2\,\delta^{(2)} (\q-\q') \frac{\rmd \sigma_\text{el}}{\rmd^2\q},
\eeq
where $\rmd \sigma_\text{el}/\rmd^2\q\simeq g^4/\q^4$ is the small angle 2 to 2 elastic cross-section, and $n$ is the density of medium color charges which is assumed to be time-independent for now.

The soft radiation spectrum can be extracted from the soft gluon radiation cross-section associated with the quark, $E \gg \omega$, as
\beq
\label{eq:spectrum-factorization}
\frac{\rmd \sigma }{\rmd \omega \rmd \Omega_p } \simeq   \frac{\rmd I}{\rmd \omega} \, \frac{\rmd \sigma_\text{vac} }{\rmd \Omega_p}  \,,
\eeq
where we notice that the Born cross-section of quark production, $\rmd \sigma_\text{vac}/ \rmd \Omega_p  =  n_f \sum_{s}  |\Mc^{(i)}_{(s)}(p)|^2$, factorizes. We note that on the right-hand-side of (\ref{eq:spectrum-factorization}) $p^+ \equiv E + \omega$.

%%%%%%%%%%%%%%%%%%%%%%%%%%%%%
\begin{figure}[t!]
\begin{center}
\includegraphics[width=0.5\textwidth]{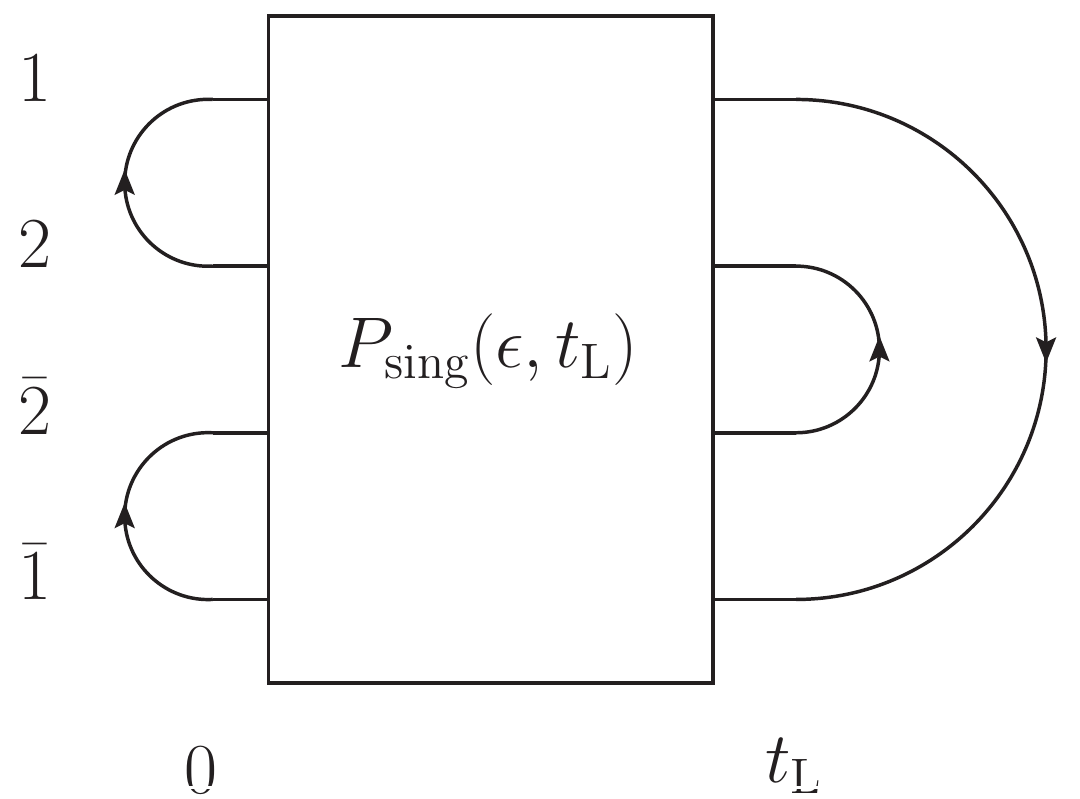} 
\caption{Real (two leftmost) and virtual (two rightmost) contributions to the emission spectrum. }
\label{fig10}
\end{center}
\end{figure}
%%%%%%%%%%%%%%%%%%%%%%%%%%%%%

Let us now discuss briefly the effect of the medium average on the soft emission spectrum.
As a consequence of the sum over color indices in the final state together with medium average, the color index $j$ of the emitter in the amplitude equals that in the complex conjugate amplitude $\bar j$  and the Wilson lines combine as follows,
\beq\label{eq:V-algebra}
\delta^{\bar j j }\big[    V^\dag (\tf, 0)  \tmat^a V^\dag (L, \tf)\big]_{\bar j i }   \big[  V(L,\ti)    \tmat^b V(\ti, 0)  \big]_{ij}  = \rmtr \big(    V^\dag (\tf, \ti) \tmat^a V(\tf,\ti)  \tmat^b\,   \big)\,,
\eeq
where we assumed that the gluon emission time in the amplitude to be larger than that in the complex conjugate amplitude, i.e.   $\tf > \ti$. A further simplification occurs when applying the Fierz identity 
\beq
\label{eq:fierz-identity-wilsonl}
\big[ V^\dag(\tf,\ti)  \tmat^a V(\tf,\ti) \big]_{ij}=   \tmat^c_{ij} \, U^{\dag ca}(\tf,\ti) \,,
\eeq
where $U(\tf,\ti) \equiv U_{_A}\big(\tf,\ti; [\x] \big)$ is a Wilson line in the adjoint representation. Using the fact that $ \rmtr  (\tmat^a  \tmat^c) = \delta^{ac}/2$, \eqn{eq:V-algebra} yields
\beq
\rmtr \big[    V^\dag(\tf,\ti)  \tmat^a   V(\tf,\ti)  \tmat^b   \big] =  \frac{1}{2}  U^{ab}(\tf,\ti) \,.
\eeq
Furthermore, we can make use of the following property of the gluon in-medium propagator,
\beq
\Gc_{_A}^{ab} (\k,L; \k', t) = \int_{\x'} \, \Gc_{_A}^{ac} (\k,L; \x', t')\,\Gc_{_A}^{cb} (\x',t'; \k', t) \,,
\eeq
where $L< t' < t$,
in order to separate out the dynamics acting between the emission time in the complex-conjugate amplitude and the end of the medium. It is then possible to show that
\beq
\label{eq:kt-intergral}
\int_\k \, \Gc_{_A}^{ab} (\k,L; \x', t' )\,\Gc_{_A}^{\dag c a} (\k,L; \y, t' )   = \delta^{(2)}(\x' -\y) \, \delta^{b c} \,,
\eeq
where $\int_\k \equiv\int \rmd^2\k/(2\pi)^2$.

Hence, integrating out the transverse momentum forces the gluon, that was {\it emitted} in the amplitude at time $t$ and transverse position $\x$, to be {\it absorbed} at time $t'$ and transverse position $\y$ in the complex conjugate amplitude as depicted in \fign{fig10}. 
Enforcing kinematical constraints in \eqn{eq:kt-intergral} leads to a more involved structure of the spectrum \cite{Wiedemann:2000tf,Salgado:2003gb}.

Anticipating our application of these techniques to multiple projectiles in Sec.~\ref{sec:derivation}, we 
presently introduce the building block 
\beq
\label{eq:W-ij}
\left.\Wc(\x_j,\tf;\x_i,\ti ) \equiv  \frac{1}{\omega^2}  (\bdel_{x}- i \omega \n_j)\cdot (\bdel_{y}+i\omega \n_i)  \, \Gc(\x,\tf ;\y, \ti) \,\rme^{-i \frac{\omega}{2}  \n_j^2 \,\tf  + i \frac{\omega}{2} \n_i^2\, \ti} \right\vert_{\substack{\y=\x_i(\ti) \\ \x=\x_j(\tf) } } \,,
\eeq
where $\n_i \equiv \p_i/E_i$ and $\x_i(t) = \n_i t$ and we have suppressed the explicit dependence on the gluon energy $\omega$. Often we will use the following shorthand $\Wc(\x_j,\tf;\x_i,\ti )\equiv \Wc_{ij}$. It describes the propagation of a gluon between the emission at position $\x_i(\ti)$ and absorption at position $\x_j(\tf)$. Note that the emission (absorption) can occur both in the amplitude or in its complex conjugate, due to the properties of the quark-gluon vertex described in \ref{sec:FeynmanRules}.

The contribution to the spectrum, depicted by the leftmost diagram in \fign{fig10}, reads then
\beq
\label{eq:bdmps-1}
\left. \omega \frac{\rmd I }{\rmd \omega } \right|_a= \frac{\alpha_s }{ 2 N_c }  \int_0^L \rmd \tf \int_0^{\tf} \rmd \ti \,   \langle    \rmTr \, U^\dag_1(\tf,\ti) \,\Wc_{11}(\tf,\ti) \rangle \,,
\eeq
where $\alpha_s = g^2/4\pi$ and we have labeled the direction of the quark with the index ``1'' ($\n_1 \equiv \n$).
The medium average of the two-point function is performed in \ref{sec:n-point-fct}, where \eqn{eq:2-point-fct} yields
\begin{align}
&\frac{1}{N_c^2-1}\langle \rmTr \,U^\dag_1\, \Wc(\x,\tf; \y,\ti) \rangle \nn
&= \frac{1}{\omega^2}(\bdel_{x}- i \omega \n_1) \cdot (\bdel_{y} + i\omega\n_1)
\tilde S^{(3)}(\x-\x_1(\tf),\y-\x_1(\ti),\0)  
\left. \rme^{ i \omega \n_1 \cdot   (\x-\y-\Delta\x_1)} \right|_{\substack{\y = \x_1(\ti) \\ \x = \x_1(\tf)} } \nn
& = 
\frac{1}{\omega^2} \bdel_{x} \cdot \bdel_{y} \, \left. \tilde S^{(3)}(\x,\y,\0) \right|_{\x=\y = 0} \,,
\end{align}
where $\Delta \x_1 = \x_1(\tf) - \x_1(\ti)$ and the 3-point function $\tilde S^{(3)}(\x,\y,\v)$ is explicitly given in \eqn{eq:3-point-fin}. For a vanishing dipole size, i.e.  $\v=\0$, we introduce the common notation,
\beq\label{eq:K-S3}
\tilde S^{(3)}(\x,\y,\0)\equiv {\cal K}(\x,\y)\,.
\eeq
Thus, the contribution from the first time-ordering reads,
\beq 
\left. \frac{\dd I}{\dd \omega}\right|_a = \frac{\alpha_s C_{F}}{\omega^3} \int_{0}^L \rmd \tf \int_{0}^{\tf}\rmd \ti \,  \bdel_{x}\cdot\bdel_{y}\, {\cal K}(\x,\y)\Big|_{\x=\y=\0} \,.
\eeq
The independence of the medium average of the position of the emitter reflects the  2D Galilean invariance of the matrix element.
Similarly, the contribution from diagram (b) in \fign{fig10} corresponds to the opposite time-ordering of emissions, namely emissions from the complex conjugate amplitude first, gives a similar result with the substitution $\ti \leftrightarrow \tf$ and $\tilde S^{(3)} \to \tilde S^{(3)\ast}$.
Adding up contributions a and b yields the spectrum of real emissions
\beq
\label{eq:bdmps-spectrum-general}
 \frac{\dd I}{\dd \omega} \equiv \left. \omega \frac{\dd I}{\dd \omega}\right|_{a+b} = \frac{\alpha_s C_{F}}{\omega^3} 2 \rmR \int_{0}^L \rmd \tf \int_{0}^{\tf}\rmd \ti \,  \bdel_{x}\cdot\bdel_{y}\, {\cal K}(\x,\y)\Big|_{\x=\y=\0} \,.
\eeq 
This spectrum was first derived by BDMPS-Z \cite{Baier:1994bd,Baier:1996sk,Baier:1996kr,Baier:1998yf,Zakharov:1996fv,Zakharov:1997uu}, see also \cite{Wiedemann:2000za,Gyulassy:2000er,Arnold:2002ja} for equivalent formulations.\footnote{Here, a note on the limits of the integrals in \eqn{eq:bdmps-spectrum-general} is in place. Note that we can safely let the upper limit of the time integral to go to $\infty$, while maintaining a finite support for the medium potential. Such a procedure would allow us to pick up finite-size effects since we would explicitly include interferences between emissions inside and outside the medium. The full BDMPS-Z spectrum accounts for these finite-size effects. Additionally, we would recover the vacuum for the piece where both $t,t' > L$. In what follows we neglect these corrections and focus on what is usually referred to as the deep LPM regime.}

Similarly, the contributions from diagrams (c) and (d) in \fign{fig10}, corresponding to virtual corrections, simply read 
\beq
\left.\frac{\dd I }{\dd \omega} \right\vert_{c+d} = - \delta(\omega) \int \dd \omega' \frac{\dd I }{\dd \omega'} ,
\eeq
as expected. This ensures that real and virtual contributions cancel for inclusive quantities, i.e., $\int \, \dd I \big|_{a+b+c+d} = 0$.

%%%%%%%%%%%%%%%%%%%%%%%%%%%%%
\begin{figure}[t]
\begin{center}
\includegraphics[width=12cm]{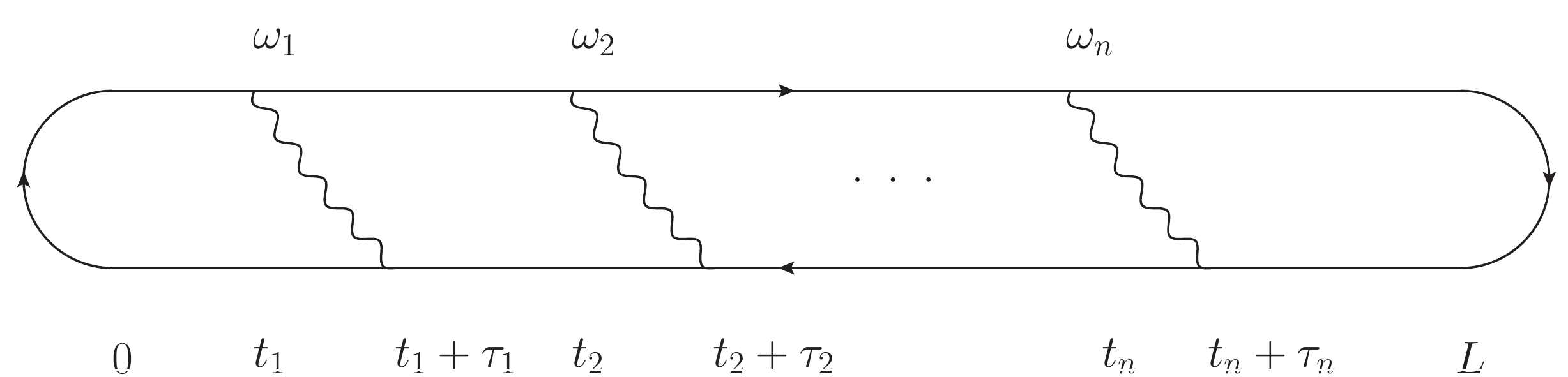}
\caption{Multiple soft gluon emissions off a quark. For each emission, $\tau_i$ is very short, and one can treat the emissions as quasi-instantaneous. Virtual diagrams are omitted in this illustration.}
\label{fig2}
\end{center}
\end{figure}
%%%%%%%%%%%%%%%%%%%%%%%%%%%%%

The expression in \eqn{eq:bdmps-spectrum-general} can be further simplified by noticing that the dominant contribution involves a strong correlation of the time-integrations.
Introducing the variable $\tau=\tf- \ti$, 
we note that its range is bounded by the coherence time $\tform \sim \sqrt{\omega/\hat q}$,
which for soft emissions $\tau \,  < \, \tform \, \ll L-\ti\,  \sim\,  L  $. Hence, in the limit of large medium one can approximate the time integration over $\tau$ as follows,
\beq
\label{eq:time-approximations}
\int_{\ti}^L \rmd \tf = \int_{0}^{L-\ti} \rmd \tau \approx  \int_{0}^{\infty} \rmd \tau \,.
\eeq
Formally, this allows to treat multiple radiation as independent, see \fign{fig2}, 
with a constant rate 
\beq
\label{eq:spect-general-2}
 \frac{\rmd I }{\rmd \omega \,\rmd t}\approx \frac{\alpha_s C_{F}}{\omega^3} \,2 \rmR \int_{0}^\infty \rmd \tau \,  \bdel_{x}\cdot\bdel_{y}\, {\cal K}(\x,\y)\Big|_{\x=\y=\0}  \,,
\eeq
where the 2-point function lives in the time interval $[t+\tau,t]$. Note that letting $\x=\y=\0$ before integrating over $\tau$ in \eqn{eq:spect-general-2} generates a spurious $\sim \tau^{-2}$ divergence that is regulated by integrating over $\tau$ before integrating over the soft gluon transverse momentum that yields the condition $\x=\y=\0$, see \cite{Blaizot:2012fh}.

In order to proceed, we must specify what we mean by the medium interaction potential. As stated earlier, we will currently only account for diffusive broadening in the plasma, corresponding to the harmonic oscillator approximation, \eqn{eq:harmonic-approx}.
We find the rate of direct emissions to be given by
\beq
\label{eq:direct-analytic}
\frac{\dd I }{\dd \omega \dd t} = \frac{\alpha_s C_F}{\pi} \sqrt{\frac{\hat q}{\omega^3}} \,,
\eeq
where we used the results in \ref{sec:harmonic-approx}, in particular \eqn{eq:ho-int}, and \eqn{eq:deriv-HO} at $\n_{12}\to 0$. Indeed, we recover the well-kown LPM rate, as in \eqn{eq:LPM-rate}.

%%%%%%%%%%%%%%%%%%%%%%%%%%%%%
\begin{figure}[t]
\begin{center}
\includegraphics[width=14cm]{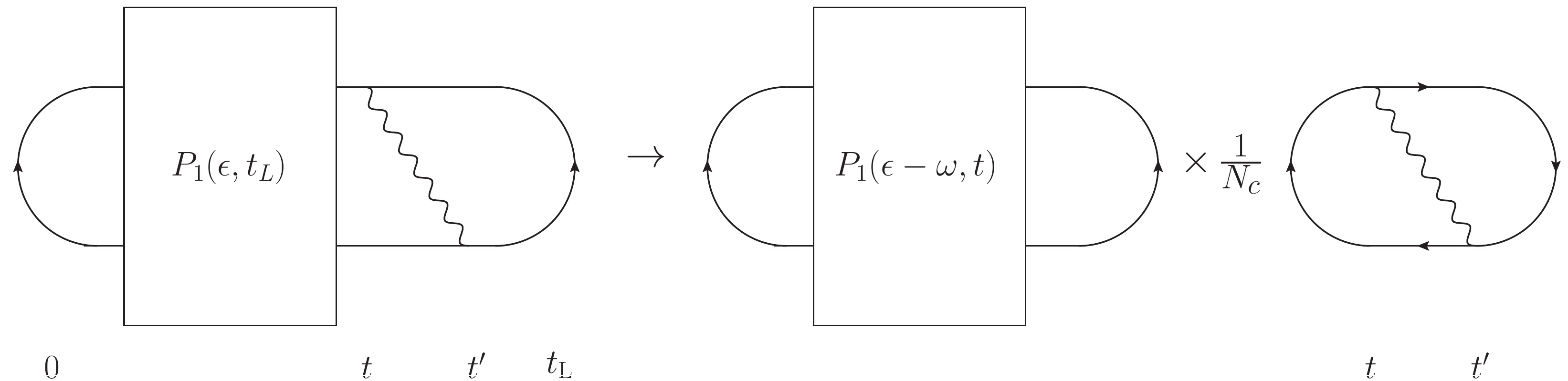}
\caption{Example diagram for the evolution of the one-prong energy loss probability $\ProbOne(\epsilon,\tend)$.}
\label{fig2-2}
\end{center}
\end{figure}
%%%%%%%%%%%%%%%%%%%%%%%%%%%%%

It is worth to highlight the probabilistic nature of multiple emissions that arise in this case. Since, in the small formation time limit $\tform \ll L$, we can treat them as independent, we easily realize that the spectrum for radiating $n$ gluons reads
\begin{align}
\label{eq:ngluon}
\frac{\rmd I^{(n)}}{\rmd \omega_1 \ldots \rmd \omega_n} &= \int_0^L \rmd t_n   \ldots  \int_0^{t_3} \rmd t_{2}\int_0^{t_2} \rmd t_{1}  \, \frac{\rmd I}{\rmd \omega_n\rmd t_n}  \ldots  \frac{\rmd I}{\rmd \omega_2\rmd t_2}  \frac{\rmd I}{\rmd \omega_1\rmd t_1} \nn
  &= \frac{1}{n! } \, \prod_{i=1}^n \int_0^L\rmd t \, \frac{\rmd I}{\rmd \omega_i \rmd t } \,.
\end{align}
This leads directly to defining an energy loss probability in the form of a Poisson distribution, as in \eqn{eq:quenching-weight-1}, that was first introduced in Ref. \cite{Baier:2001yt}, see also \cite{Arleo:2002kh,Salgado:2003gb,Baier:2006fr}.  However, in line with the derivations of Sec.~\ref{sec:derivation}, it is convenient to use an alternative way of expressing this probability in terms of a rate equation. As depicted in \fign{fig2-2}, one of the corrections to the probability at time $\tend$ reads simply
\beq 
\label{eq:1prong-rate-equation-detail}
\Delta \ProbOne (\epsilon, \tend)\Big|_a =\alpha_s C_F \,2 \rmR \int_{0}^\tend \rmd \ti  \int_{\ti}^{\tend}\rmd \tf \int_0^\infty   \frac{\rmd \omega }{\omega^3}\,  \bdel_{x}\cdot\bdel_{y}\,{\cal K}(\x,\y)\left.\right\vert_{\x=\y=\0} \,  \ProbOne(\epsilon-\omega, \ti) \,,
\eeq
see \eqn{eq:bdmps-spectrum-general}.
To obtain the full correction we should sum the contribution from all the diagrams in \fign{fig10}, and we will also perform the same set of approximations as described in \eqn{eq:time-approximations}, in particular regarding the branching time. The total correction then reads
\beq
\Delta \ProbOne (\epsilon, \tend)\Big|_{a+b+c+d} = \int_{0}^\tend \rmd \ti  \int_0^\infty \rmd \omega \, \left[ \frac{\rmd I}{\rmd \omega\, \dd t} -\delta(\omega)\int_0^\infty  \rmd \omega' \frac{\rmd I}{\rmd \omega'\, \dd t}\right]  \,  \ProbOne(\epsilon-\omega, \ti) \,.
\eeq
Hence, the total probability obeys the following equation
\beq
\ProbOne(\epsilon,\tend) = \delta(\epsilon) + \Delta \ProbOne (\epsilon, \tend)\Big|_{a+b+c+d} \,,
\eeq
where the first term corresponds to the absence of energy loss. By taking a derivative with respect to the final time, we obtain the following evolution equation for the energy-loss probability,
\beq
\label{eq:eloss-rate-eq}
\frac{\del }{\del t} \ProbOne(\epsilon, t ) = \int_0^\infty  \rmd \omega \, \Gamma(\omega,t)\ProbOne(\epsilon-\omega, t ) \,,
\eeq
where we have used the notation from \eqn{eq:gamma-ij}, with $\Gamma(\omega,t) \equiv \Gamma_{11}(\omega,t)$.
It is a straightforward exercise to check, that we also obtain this equation by acting with a time derivative directly on \eqn{eq:quenching-weight-1}.

The solution to \eqn{eq:eloss-rate-eq}, or equivalently \eqn{eq:quenching-weight-1}, can be found using a Laplace transform of the energy loss probability,
\beq
\ProbOne(\epsilon,L) = \int_C \frac{\dd \nu}{2\pi i} \, \tilde \ProbOne(\nu,L) \rme^{\nu \epsilon} \,,
\eeq
where the contour $C$ runs parallel to the imaginary axis to the right of any singularity of $\tilde \ProbOne(\nu,L)$ in the complex-$\nu$ plane. 
Inserting this into the evolution equation, we get
\beq
\label{eq:1prong-evolution-equation-Mellin}
\frac{\del }{\del t} \tilde \ProbOne(\nu, t ) = \gamma(\nu,t) \tilde \ProbOne(\nu, t ) \,,
\eeq
where $\gamma(\nu,t) = \int_0^\infty \dd \omega \, \Gamma(\omega,t) \rme^{-\nu \omega} = -2\sqrt{\pi \nu \, \bar \alpha^2 \hat q}$
is the Laplace transform of the (regularized) splitting rate. 
The solution to \eqn{eq:1prong-evolution-equation-Mellin}, with initial condition $\tilde \ProbOne(\nu,0) = 1$, is simply
\beq
\label{eq:ProbOneMellin}
\tilde \ProbOne(\nu,L) 
= \rme^{\int_0^L \dd t \, \gamma(\nu,t)} = \rme^{- 2\sqrt{ \pi \nu \, \bar \alpha^2 \hat q}L } \,.
\eeq
Taking the inverse Laplace transform, we recover the simple form of the energy loss probability, given in \eqn{eq:one-prong-eloss} \cite{Baier:2001yt}.

%%%%%%%%%%%%%%%%%%%%%%%%%%%%%
\section{Two-parton energy loss and derivation of \eqn{eq:main-result}}
\label{sec:derivation}
%%%%%%%%%%%%%%%%%%%%%%%%%%%%%

%%%%%%%%%%%%%%%%%%%%%%%%%%%%%
\begin{figure}[t!]
\begin{center}
\includegraphics[width=0.4\textwidth]{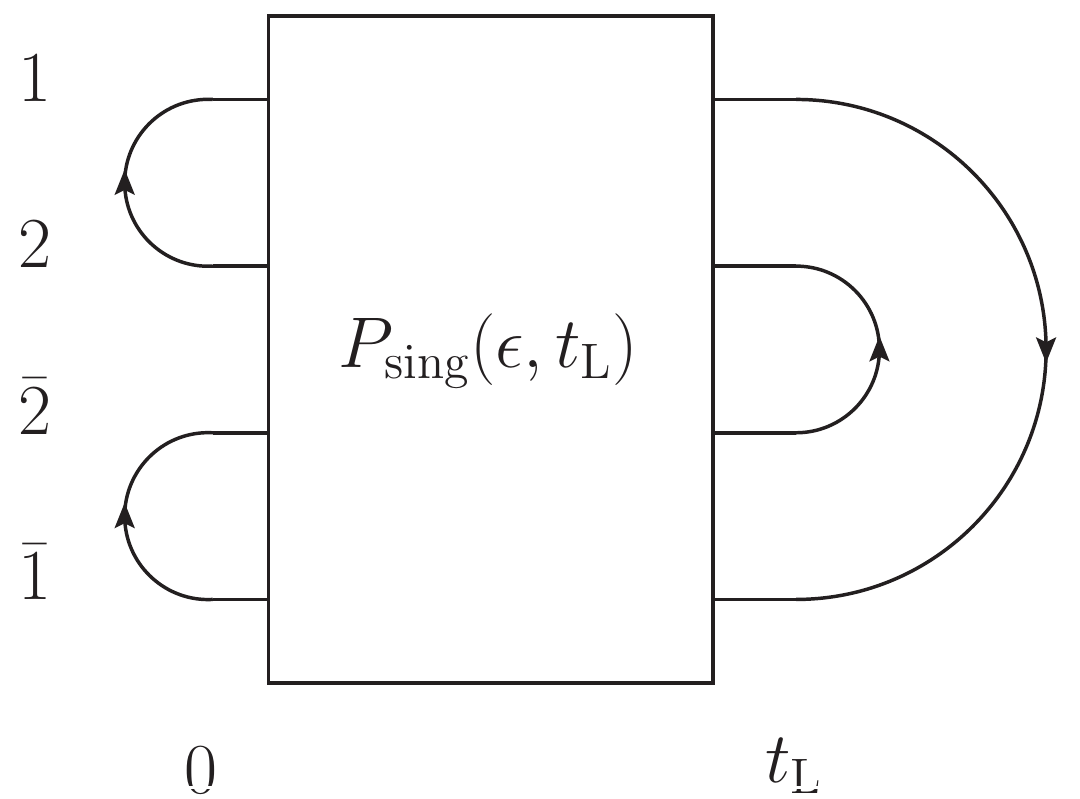}
\caption{The 2-prong energy-loss probability. The relation to the Feynman diagrams in \fign{fig5} and \fign{fig7} is obvious, and is obtained by deforming the Wilson lines in the complex conjugate amplitude to appear below the ones in the amplitude.}
\label{fig8}
\end{center}
\end{figure}
%%%%%%%%%%%%%%%%%%%%%%%%%%%%%

In Sec.~\ref{sec:two-prong-solution}, we anticipated the final result for the two-pronged energy loss distributions that is shown to take a rather simple form. For completeness, we shall in the following present a complete derivation of \eqn{eq:main-result}. 

Recall that all medium averages are approximated by the harmonic approximation, we refer to \ref{sec:harmonic-approx} for detailed calculations of all relevant two- and three-point functions.
We shall proceed by constructing an evolution equation for the two-prong probability of a color singlet dipole, depicted in \fign{fig8}, that resums multiple emissions, including for the first time interferences and virtual contributions between the two prongs. 
We note that at zeroth order, that is in the absence of radiation, the energy loss probability must be given by $\delta(\epsilon)$. Then, as within the Dyson-Schwinger construction, we will evaluate the correction to the energy-loss probability at a late time, where the building block in the intermediate evolution is the color matrix 
$\Mmat^{i j \bar j \bar i}_{kl \bar l \bar k}(\omega,t) $ which evolves the set of lower color indices ($\{k,l,\bar l,\bar k\}$ associated with legs $12\bar 2 \bar 1$, respectively) at the initial time to the set of upper indices ($\{i,j,\bar j,\bar i \}$) at time $t$, keeping track of the amount of energy $\omega$ that has been lost through medium-induced emissions in the process.
The two-prong probability is defined as the (diagonal) projection,
\beq
\label{eq:PsingNorm}
\ProbSing(\omega,t) \equiv \frac{1}{N_c} \delta_{i\bar i} \delta_{j \bar j} \Mmat^{ij \bar j \bar i}_{kl \bar l \bar k}(\omega,t) \delta^{kl} \delta^{\bar l \bar k} \,,
\eeq
see \fign{fig8}, 
where we have projected on the initial state and the normalization factor $N_c^{-1}$ ensures that $\ProbSing(\omega,t) = \delta(\omega)$ in the absence of medium-induced {\sl emissions}.\footnote{According to this notation, when there are no emissions, the color structure reduces to $\Mmat^{ijji}_{kkll} \sim \langle \rmtr (V_2^\dag V_2 V_1^\dag V_1) \rangle \sim N_c$. The normalization is therefore equivalent to the averaging over quark color states.}
The singlet two-pronged quenching weight is then given by $\ProbSing(\epsilon,\tend) = \delta(\epsilon) + \Delta \ProbTwo(\epsilon,\tend)$, where the correction factor contains the higher-order effects from radiation. From the locality of the medium averages we can write this correction factor as
\beq
\label{eq:PsingMastEq}
\Delta \ProbSing(\epsilon, \tend)  = \alpha_s \int_0^\tend \dd \ti \int_\ti^\tend \dd \tf \, \Rmat(\tf,\ti) \otimes \Mmat(\ti) + \ldots \,,
\eeq
where the ellipses imply an equivalent term with the opposite time-ordering, $\Rmat$ is a {\it radiation} matrix that acts in the time interval $[\tf,\ti]$ between emission in the amplitude and complex-conjugate amplitude. We have defined $\alpha_s \equiv  g^2\big/4\pi$. The evolution matrix is implicitly projected onto the proper initial state ensuring that the expression is properly normalized, according to \eqn{eq:PsingNorm}. The convolution is defined by
\beq
\Rmat(\tf,\ti) \otimes \Mmat(\ti) = \int_0^\infty \frac{\dd \omega}{\omega}\, \Rmat_{ij \bar j \bar i}(\omega,\tf,\ti) \Mmat^{ij \bar j \bar i}_{kkll}(\epsilon-\omega,\ti) \,.
\eeq
Due to the two different ways of re-connecting the final-state indices, radiation will give rise to two contributions to the rate equation. We will refer to them as direct and interference contributions. In the latter case, a set of virtual diagrams are resummed into a parameter that governs the coherence properties of the pair.
In this way, we derive a new rate equation for two-prong structures that propagate through the medium. We organize our discussion into considering direct emissions, interferences and contributions to the decoherence parameter separately.

%%%%%%%%%%%%%%%%%%%%%%%%%%%%%
\subsection{Direct emissions from the $q\bar q$ pair}
\label{sec:direct-emissions}
%%%%%%%%%%%%%%%%%%%%%%%%%%%%%

%%%%%%%%%%%%%%%%%%%%%%%%%%%%%
\begin{figure}[t!]
\begin{center}
\includegraphics[width=15cm]{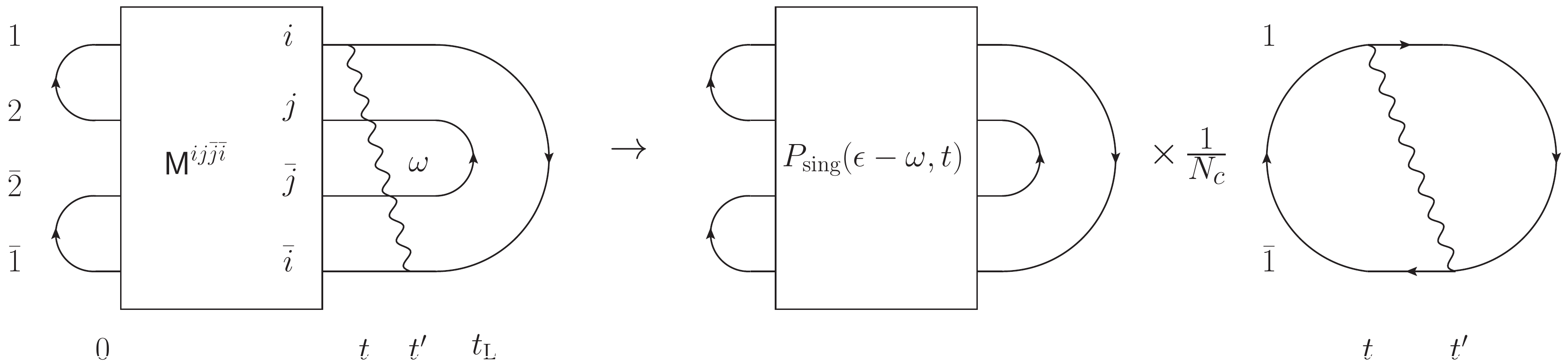}
\caption{Diagrams describing a direct emission from leg 1 of the $q \bar q$ pair.}
\label{fig9}
\end{center}
\end{figure}
%%%%%%%%%%%%%%%%%%%%%%%%%%%%%

At some late time, let us consider the direct emission of a real gluon, i.e. emitted and absorbed by the quark (e.g.  ). By late, we mean that there are no more emissions between the absorption time and the evaluation time $\tend$. The emission couples to a matrix element, see diagram on the left hand side in \fign{fig9}. 
From this figure we read off the radiation matrix for direct emissions,
\begin{align}
\Rmat^{\rm dir}_{ij \bar j \bar i} (\tf,\ti) &= \frac{1}{N_c} \llangle \big[V_{1}^\dag(\tf, \ti) \tmat^b V_{1}^\dag(\tend, \tf) V_ 1(\tend ,\ti) \tmat^a \big]_{\bar i i}  \, \big[  V_2( \tend, \ti) V^\dag_{2}(\tend, \ti ) \big]_{ j \bar j}\Wc^{ba}(\x_1, \tf;\x_1,\ti) \rrangle\nn
&=  \frac{1}{N_c}  \llangle \big[ U_{1}^\dag(\tf,\ti) \Wc(\x_1,\tf; \x_1, \ti) \big]^{ca}  \rrangle \,   \left[\tmat^c \tmat^a \right]_{\bar i i} \delta_{ j\bar j} \,,
\end{align}
where $\x_1(\ti)$ ($\x_1(\tf)$) is the position of the quark at the time of emission (absorption) and we have used the Fierz identity in Eq.~(\ref{eq:fierz-identity-wilsonl}). The generalized Green's function $\Wc^{ab}$ is defined in \eqn{eq:W-ij}.
Due to color conservation, which is ensured by the correlator \eqn{eq:med-average}, we can anticipate that, after performing the medium averaging, the correlator becomes
\beq
\llangle \big[U_{1}^\dag\Wc_{11} \big]^{ca}\rrangle = \frac{\delta^{ca}}{N_c^2-1} \langle \rmTr\, U_1^\dag\Wc_{11} \rangle \,,
\eeq
which leads to $\delta^{ca} \left[ \tmat^c \tmat^a \right]_{\bar i i} = C_F \delta_{\bar i i}$. 
Hence, $\Rmat^{\rm dir}_{ij \bar j \bar i} \sim \delta_{i \bar i} \delta_{j \bar j}$, see the two diagrams on the right side of the arrow in \fign{fig9} where it is made explicit that the correction appears with a factor $1/N_c$. This diagonalizes the matrix $\Mmat^{i j j i}(12\bar 2 \bar 1)$, thus reproducing the original color structure of $\ProbSing(\epsilon,\ti)$ at intermediate times, see \eqn{eq:PsingNorm}.
This contribution leads to a correction to the two-pronged probability that reads
\beq 
\left. \Delta \ProbSing(\epsilon, \tend )\right|^{\rm dir}_a =  \frac{\alpha_s}{2 N_c }  \int_{0}^\tend \rmd \ti \int_{\ti}^\tend \rmd \tf \int_0^\infty   \frac{\rmd \omega }{\omega}\,   \langle \rmTr \, U^\dag_1( \tf,\ti) \Wc_{11}( \tf, \ti)    \rangle \,  \ProbSing(\epsilon-\omega,\ti ) \,,
\eeq
for this particular time ordering.
The evolution kernel is identical to the one we considered for the one-particle energy loss for a particular time ordering, cf. \eqn{eq:1prong-rate-equation-detail}, and we can therefore simply refer to Sec.~\ref{sec:QuenchingWeights} for further details.
After summing real and virtual diagrams, as in \fign{fig10}, we obtain the full correction for the quark emissions. The resummation for emissions off the antiquark (leg ``2'') proceeds in a completely analogous fashion. The full correction factor from direct emissions therefore reads
\beq
\left. \Delta \ProbSing (\epsilon, \tend)\right|^{\rm dir}=   \int_{0}^\tend \rmd t  \int_0^\infty  \rmd  \omega\, \left[  \Gamma_{11}(\omega, t) +  \Gamma_{22}(\omega, t) \right]\,   \ProbSing(\epsilon-\omega,t) \,,
\eeq
where $\Gamma_{11}(\omega,t)$ is defined in Eq.~(\ref{eq:gamma-ij}) ($\Gamma_{22}(\omega,t)$ is found by a trivial change of indices).
Together these expressions simply resum independent soft gluon emissions off each leg, and constitute the limit of completely incoherent energy loss.

%%%%%%%%%%%%%%%%%%%%%%%%%%%%%
\subsection{Interference emissions off the $q\bar q$ pair}
\label{sec:interference}
%%%%%%%%%%%%%%%%%%%%%%%%%%%%%

%%%%%%%%%%%%%%%%%%%%%%%%%%%%%
\begin{figure}[t!]
\begin{center}
\includegraphics[width=15cm]{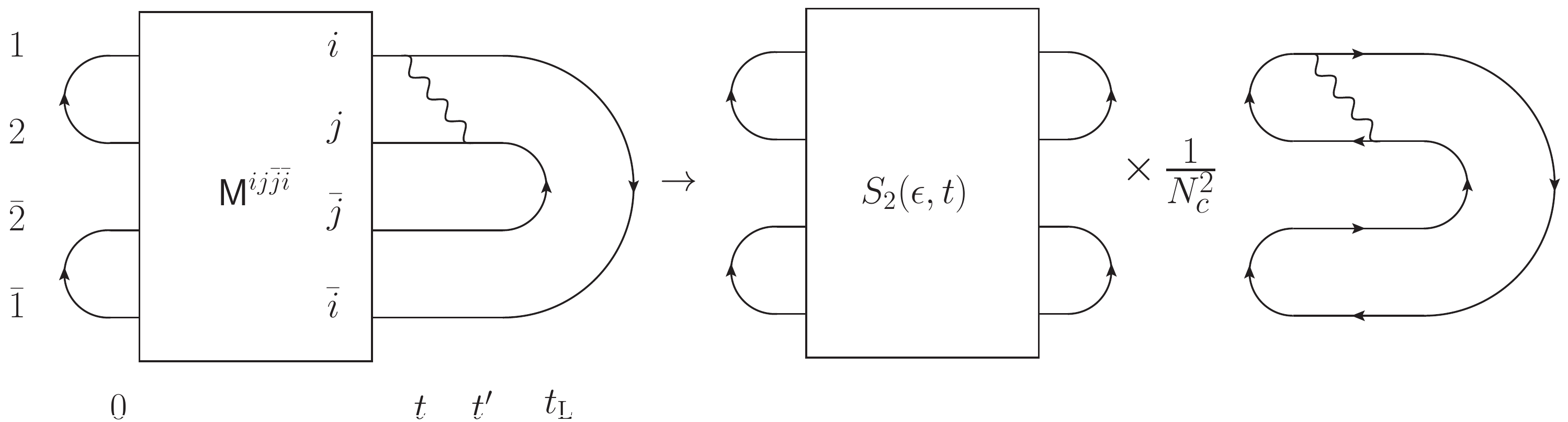} 
\caption{ The interference (flip) contributions. }
\label{fig12}
\end{center}
\end{figure}
%%%%%%%%%%%%%%%%%%%%%%%%%%%%%

Let us now turn our attention to the interference contributions,
considering first the first diagram in \fign{fig12}.
In this case, the correction term involves the radiation matrix
\begin{align}
\label{eq:flip-V-prod}
\Rmat^{\rm int}_{ij \bar j \bar i}(\tf,\ti)&= \frac{1}{N_c} \llangle \big[ V_{1}^\dag(\tend , \ti) V_ 1(\tend, \ti ) \tmat^a \big]_{\bar i i}  \, \big[V^\dag_ 2( \tend, \ti) V_ 2(\tend, \tf) \tmat^b V_ 2( \tf, \ti) \big]_{j \bar j} \Wc^{ba}(\x_2, \tf; \x_1,\ti ) \rrangle \nn
&= \frac{1}{N_c} \llangle \big[ U^\dag_2( \tf, \ti) \Wc(\x_2, \tf; \x_1, \ti) \big]^{ca} \rrangle \, \tmat^a_{\bar i i} \tmat^c_{ j \bar j}  \,.
\end{align}
Once again, we can anticipate the relevant color structure based on the fact that we are left with two Wilson lines in the adjoint representation whose average generates a delta function in color space,
\beq
\llangle \big[ U^\dag_2 \Wc_{21} \big]^{ca}\rrangle = \frac{\delta^{ca}}{N_c^2-1} \langle \rmTr \,U^\dag_2  \, \Wc_{21} \rangle \,.
\eeq
Using the Fierz identity, that simplifies $\tmat^a_{\bar i i}\tmat^a_{j \bar j} = (\delta_{\bar i \bar j} \delta_{ij} - \delta_{\bar i i}\delta_{\bar j j}/N_c)/2$,we can show that
\beq
\label{eq:P2-flip-1}
\left. \Delta \ProbSing(\epsilon, \tend )\right|^{\rm int}_a = -  \frac{\alpha_s}{2 N_c}  \int_{0}^\tend \rmd \ti \int_{\ti}^\tend \rmd \tf \int_0^\infty   \frac{\rmd \omega }{\omega}\,   \langle \rmTr \, U^\dag_2( \tf,\ti) \Wc_{21}( \tf, \ti)\rangle \, S_2(\epsilon-\omega,\ti) \,,
\eeq
where the negative sign arises from the product of vertices and $S_2$ is defined as
\beq
\label{eq:s2def}
S_2(\omega,t) \equiv \frac{1}{N_c^2-1} \left[\Mmat^{ii jj}_{kkll}(\omega,t) - \frac{1}{N_c} \Mmat^{ij ji}_{kkll}(\omega,t) \right] \,.
\eeq
Hence the interference contribution induces a coupling between the diagonal parts of the propagation matrix, $\Mmat^{ijji} \sim N_c \ProbSing$, with its non-diagonal components, represented by $\Mmat^{ii jj}$. We will discuss this object in more detail, in particular its convenient simplification in the large-$N_c$ limit, in Sec.~\ref{sec:radiative-corrections}.

In the time interval where the emission is taking place, we again encounter a 2-point function, connecting the quark and the antiquark. Using \eqn{eq:W-ij}  and \eqn{eq:2-point-fct}, we find that 
\begin{align}
&\frac{1}{N_c^2-1}\langle \rmTr \,U^\dag_2  \, \Wc(\x,\tf; \y,\ti) \rangle  \nn
&=\frac{1}{\omega^2}\rme^{ i \frac{\omega(\n_1^2-\n_2^2)}{2} \ti}\, (\bdel_{x}- i \omega \n_2) \cdot (\bdel_{y}+ i \omega \n_1) \tilde S^{(3)}(\x-\x_2(\tf),\y-\x_2(\ti),\0)\, \rme^{ i \omega \n_2 \cdot   (\x-\y-\Delta \x_2) } \nn
&=\frac{1}{\omega^2} \rme^{ i \frac{\omega(\n_1^2-\n_2^2)}{2} \ti + i \omega \n_2 \cdot   (\x-\y-\Delta \x_2)}\, \bdel_{x}\cdot (\bdel_{y} + i\omega \n_{12})  \tilde S^{(3)}(\x-\x_2(\tf),\y-\x_2(\ti),\0) \,,
\end{align}
where $\Delta \x_2 = \x_2(\tf)- \x_2(\ti)$. Note that, in comparison with \eqn{eq:2-point-fct}, we have shifted the coordinates with respect to the Wilson line at coordinate $\x_2$ due to the presence of $U_2$.
Finally, enforcing the boundary conditions $\y=\x_1(\ti)$ and $\x=\x_2(\tf)$, we find
\beq
 \frac{1}{N_c^2-1}\langle \rmTr \,U^\dag_2  \, \Wc(\x_2,\tf;\x_1,\ti) \rangle  = \frac{1}{\omega^2}\,  \rme^{ i \frac{\omega}{2} \n_{12}^2 \ti} \bdel_{x} \cdot (\bdel_{y}+ i\omega \n_{12}) \tilde S^{(3)}(\x,\y,\0)\Big|_{\x=\0,\y=\x_{12}} \,,
\eeq
with $\x_{12} \equiv \n_{12} t$.
Inserting the above expression into \eqn{eq:P2-flip-1}, we find that
\begin{align}
\label{eq:P2-flip-2}
\left.\Delta \ProbTwo(\epsilon, \tend ) \right|^{\rm int}_a &=-\alpha_s C_F\int \frac{\rmd \omega }{\omega^3 } \int_0^\tend \rmd \ti \int_{\ti}^\tend \rmd \tf \,\rme^{ i \frac{\omega}{2} \n_{12}^2 \ti} \bdel_{x} \cdot (\bdel_{y}+ i\omega \n_{12}) {\cal K}(\x,\y)\Big|_{\x=\0,\y=\x_{12}} \nn
&\times S_2(\epsilon-\omega,\ti) \,,
\end{align}
where we denoted $\tilde S^{(3)}(\x,\y,0) \equiv {\cal K}(\x,\y)$.
Furthermore, accounting for all topologies (as in \fign{fig10}, where the upper leg is denoted ``1'', and has support in the amplitude, and the lower leg ``$\bar 2$'' that has support in the c.c. amplitude), yields 
\beq
\left. \Delta \ProbSing(\epsilon, \tend ) \right|^{\rm int}  = \int_0^\tend \dd \ti \int_0^\infty \rmd \omega \, \left[ \Gamma_{21}(\omega,t) + \Gamma_{12}(\omega,t) \right] S_2(\epsilon-\omega, \ti), 
\eeq
where we have applied the approximation \eqn{eq:time-approximations} and $\Gamma_{ij}(\omega,t)$ is defined according to \eqn{eq:gamma-ij} with\footnote{It is understood throughout the paper that the vacuum part is implicitly subtracted $\mathcal{K} (\x,\y) \to \mathcal{K}(\x,\y)-\mathcal{K}_0(\x,\y)$ since we are only interested in the medium-induced contribution. This subtraction is formally necessary in order to recover that $\lim_{\theta_{12}^2 \to 0}\dd I_{21} = - \dd I_{11}$, as should be.} 
\beq
\label{eq:antenna-spect}
\frac{\rmd I_{21}}{\rmd \omega\, \rmd \ti} =-\frac{\alpha_s C_F}{\omega^3}  2\rmR\,  \int_0^\infty \rmd \tau\,  \rme^{ i \frac{\omega}{2} \n_{12}^2 \ti }\, (\bdel_{y}+i\omega \n_{12})\cdot \bdel_{x} \, {\cal K}(\x,\y)\left.\right\vert_{\x=\0,\y=\x_{12}} \,,
\eeq
and, analogously, $\dd I_{12} \big/ (\dd \omega \, \dd t)$ with $\x \leftrightarrow \y$. However, note that 
%\old{$\tilde S^{(3)}(\x,\y,\v) =\tilde S^{(3)}(\y,\x,\v)$ is symmetric in the two first arguments}
${\cal K}(\x,\y)$ is symmetric in its arguments, see \eqn{eq:ho-int}, and therefore $\dd I_{21} = \dd I_{12}$.
We recover with \eqn{eq:antenna-spect} the leading order antenna radiation spectrum \cite{CasalderreySolana:2011rz,MehtarTani:2011jw,MehtarTani:2012cy}. Finally, in the harmonic oscillator approximation, \eqn{eq:harmonic-approx}, we find that
\begin{align}
\label{eq:deriv-HO}
(\bdel_{y} + i\omega \n_{12})\cdot\bdel_{x}\,\left. \mathcal{K}(\x,\y) \right\vert_{ \x=0,\y =\x_{12}} &= -\frac{1}{2\pi} \left(\frac{\omega \Omega}{\sinh \Omega \tau} \right)^2 \rme^{i\frac{\omega \Omega}{2}\x_{12}^2 \coth \Omega \tau } \nn
&\times \big(2 + i\omega \n_{12} \cdot \x_{12}+ i\omega \Omega\x_{12}^2 \coth \Omega \tau \big) \,,
\end{align}
where $\Omega = (1+i)\sqrt{\hat q /\omega}/2$.  

In addition to the formation time of the medium-induced radiation $\tform \sim \sqrt{\omega/\hat q }$, the above interference spectrum depends explicitly on two other time scales,  
\beq 
t_\text{quant} \sim \frac{1}{\omega\theta_{12}^2} \,, \quad  \text{and}  \quad t_\text{res}\sim  \frac{1}{\theta_{12}(\hat q \omega )^{1/4}}.
\eeq 
The former is encoded in the phase appearing in \eqn{eq:antenna-spect}  and pertains to the loss of coherence due to quantum decoherence of the pair.
Since this time-scale does not depend on any medium parameter, we refer to it as the quantum-mechanical decoherence time.
The latter timescale which can be read off the phase in \eqn{eq:deriv-HO}, is related to the suppression of interferences when the medium-induced radiation resolves the the pair, i.e., $k_\text{f}^{-1} \ll |\x_{12}|$, where $k_\text{f} = \omega \theta_\text{f}$ is the accumulated momentum at formation time. In terms of the color decoherence time, we can also write
\beq
t_\text{quant} \sim \left( \frac{\theta_\text{f}}{\theta_{12}}\right)^{4/3} \tdecoh \,, \quad  \text{and} \quad \tres \sim  \left( \frac{\theta_\text{f}}{\theta_{12}}\right)^{1/3} \tdecoh \,.
\eeq
Since soft, large-angle radiation governs energy-loss we will mainly be interested in the regime $\theta_\text{f} \gg \theta_{12}$.
This leads to the following ordering of timescales: $\tdecoh \ll \tres \ll \tquant$.

The ordering of these three time-scales is reversed when considering small angle radiation, i.e. $\theta_\text{f} < \theta_{12}$ leads to $\tquant < \tres < \tdecoh$. This corresponds to small-angle and therefore hard radiation that does not contribute to the energy-loss of the composite, two-prong system. In this regime, the interferences are suppressed in this regime leading to independent emissions off the two dipole constituents, see \eqn{eq:direct-analytic}. However, we argue that these also correspond to rare emissions, as long as the soft scale that governs energy loss corresponds to parametrically larger angles, $\abar^2 \omega_c \ll (\hat q /\theta_{12}^4)^{1/3} < \omega_c$. Leaving a more careful treatment of this regime for future studies, for the applications considered in the present paper, namely energy-loss of multi-parton systems, this regime can safely be neglected.

The discussion above demonstrates that we therefore can approximate the spectrum as,
\begin{align}
\label{eq:spectrum-intermediate}
\frac{\rmd I_{21}}{\rmd \omega\, \rmd \ti} 
\approx -\frac{\alpha_s C_F}{\pi} \sqrt{\frac{\hat q}{\omega^3}} \Theta\big( t - \min(\tres,\tquant) \big) \,,
\end{align}
where, for the relevant regime of large angles, $\tdecoh < \min(\tquant,\tres)$, so that the spectrum will be suppressed by color decoherence effects, see Sec.~\ref{sec:radiative-corrections}, before the condition in the theta-function becomes relevant.

%%%%%%%%%%%%%%%%%%%%%%%%%%%%%
\subsection{Radiative corrections to the decoherence parameter }
\label{sec:radiative-corrections}
%%%%%%%%%%%%%%%%%%%%%%%%%%%%%

%%%%%%%%%%%%%%%%%%%%%%%%%%%%%
\begin{figure}[t!]
\begin{center}
\includegraphics[width=0.8\textwidth]{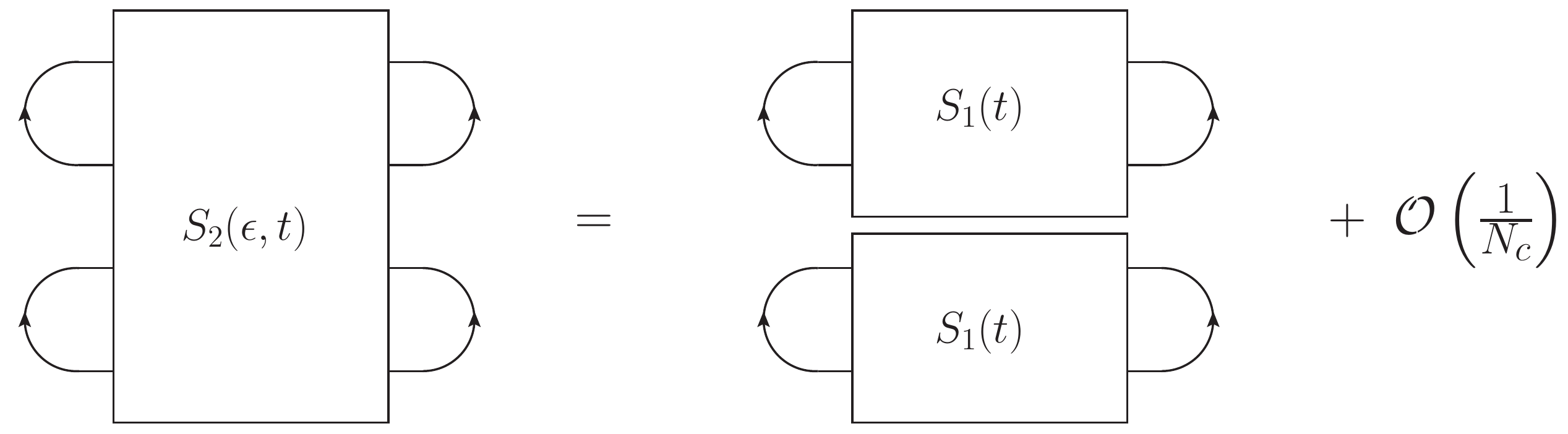} 
\caption{ The factorization of two dipoles in the large $N_c$}
\label{fig13}
\end{center}
\end{figure}
%%%%%%%%%%%%%%%%%%%%%%%%%%%%%

We are now left with evaluating radiative contributions to $S_2$, defined in \eqn{eq:s2def}, which mixed diagonal and non-diagonal terms of the evolution matrix $\Mmat^{ij \bar j \bar i}$. This couples the evolution of the singlet two-prong distribution to a hierarchy of complicated color structures and is a priori a formidable task. Surprisingly, we will find a convenient simplification that allows us to obtain a closed expression for the evolution of $\ProbSing(\epsilon,t)$.

In order to make contact with previous results, let us first consider one-gluon emission.
In the absence of further emissions the matrix structure reduce to $\Mmat^{ij ji} \sim N_c$ and $\Mmat^{ii jj} \sim \langle \rmtr(V_2^\dag V_1 ) \rmtr(V_1^\dag V_2)  \rangle$, and at finite-$N_c$, this object becomes $S_2(\omega,t) |_{\rm no rad}= \delta (\omega) S^{(0)}_2(t)$, where
\begin{align}
S^{(0)}_2(t) &= \frac{1}{N_c^2 -1} \left( \langle \rmtr(V_2^\dag V_1 ) \rmtr(V_1^\dag V_2) \rangle -1 \right) = \frac{1}{N_c^2 -1} \langle \rmTr \,U_2 U_1^\dag \rangle \,,
\end{align}
which is understood as the decoherence parameter of the pair that governs the color decoherence of the antenna, which was analyzed for the first time in \cite{MehtarTani:2011tz}. 

The evolution of $S_2$ in course of multiple emissions, in particular the $\Mmat^{iijj}$ component of the evolution matrix, couple to a hierarchy of higher-order colour structures. However, a significant simplification of the problem emerges in the large-$N_c$ limit where $S_2$ factorizes into the product of two independent dipoles \cite{Dominguez:2012ad}, 
\beq
S_2 (\omega,\ti) = \delta(\omega) S_1(\ti) S^\ast_1(\ti) + \mathcal{O} \left(\frac{1}{N_c^2} \right) \,,
\eeq
see \fign{fig13}. 
We have provided some further discussion about this crucial point in \ref{sec:before-interference}, where we argue in general terms that any radiative (real) contribution to $S_2$ are subleading in $1/N_c$  or, as depicted in \fign{fig7} (right), subleading in $\tbr/L$ due to overlapping formation time. Due to the latter point, gluons that cross the cut are subleading, and thus $S_2$ does not contribute to energy loss. It can also be shown that $S_1$ is real valued.  

%%%%%%%%%%%%%%%%%%%%%%%%%%%%%
\begin{figure}[t!]
\begin{center}
\includegraphics[width=\textwidth]{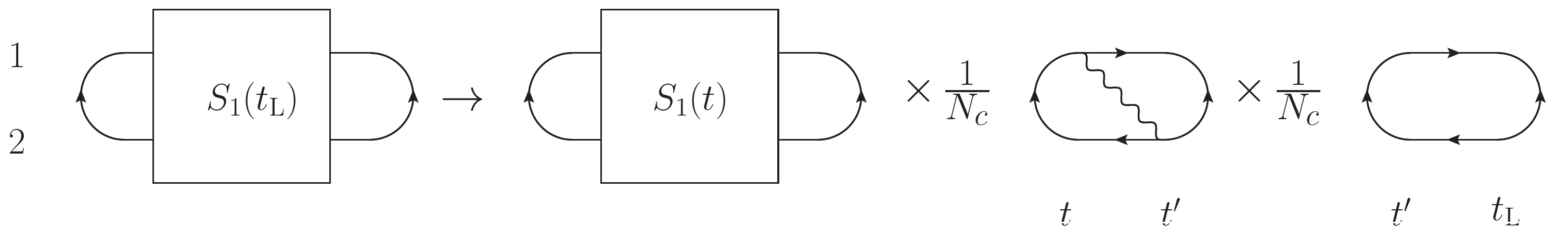} 
\caption{Single dipole evolution, see \eqn{eq:decoherence-par-evolution}.}
\label{fig14}
\end{center}
\end{figure}
%%%%%%%%%%%%%%%%%%%%%%%%%%%%%

The correction to $S_1$ due to a single emission, see \fign{fig14}, is given by
\beq
\Delta S_1(\tend,0) =   \int_0^\tend \rmd \ti \, \big\langle \rmtr \, V_2^\dag ( \tend, \ti)  V_1 (\tend,\ti) \big\rangle\,  \Sigma_{12}(\ti)\,   S_1(\ti,0) \,,
\eeq
where we recognize the dipole cross-section,
\beq
S_1^{(0)}(\tend,\ti)  = \frac{1}{N_c}\big\langle \rmtr \, V_2^\dag (\tend,\ti)  V_1 (\tend,\ti)  \big\rangle \,,
\eeq
and $\Sigma_{12}(\ti)$ denotes the exchange of a (virtual) gluon. 
In the harmonic-oscillator approximation, employed in \ref{sec:harmonic-approx}, 
we have
\beq
\label{eq:dipole-xsection}
S_1^{(0)}(\tend,0)  = \exp\left[-\frac{1}{4} \int_0^\tend \rmd t\, \hat q_{_F} \x^2_{12}(t) \right] \,,
\eeq
where $\hat q_{_F}$ is the jet quenching parameter in the fundamental representation.
This factor leads to a suppression of the interference contributions at times larger than a characteristic decoherence time, which parametrically goes as $\tdecoh \sim (\hat q  \theta_{12})^{-\onethird}$. As mentioned above, this factor will suppress the interference terms at late times.

One can also write the corresponding Dyson-Schwinger equation,
\beq
\label{eq:decoherence-par-evolution}
S_1(\tend,0) = S_1^{(0)}(\tend,0)  +   \int_0^\tend \rmd \ti \, S_1^{(0)}(\tend,\ti)  \,  \Sigma_{12}(\ti) \,  S_1(\ti,0) \,,
\eeq
which is solved by
\beq
\label{eq:decoherence-solution}
S_1(t,0)  = S_1^{(0)}(t,0) \exp\left[ \int_0^t \rmd t' \, \Sigma_{12}(t') \right] \,.
\eeq
The additional exponent in \eqn{eq:decoherence-solution} accounts for radiative corrections to the jet quenching parameter, as will become clear from the discussion below.

One of the four contribution to $\Sigma_{12} (\ti)$ can be read directly off from \fign{fig14} and reads
\beq
\Sigma_{12}(t) \Big|_a = \frac{g^2 }{8 \pi N_c}   \int_0^\infty  \rmd \tau   \int_0^ \infty \frac{\rmd \omega}{\omega}\, \big\langle  \rmtr\,  \left(V^\dag_2(\tf,\ti) \tmat^b V_1(\tf,\ti)  \tmat^a \right) \Wc^{ba}(\x_2,\tf ;\x_1, \ti) \big\rangle \,,
\eeq
where we already have employed our standard approximations in the limits of the integral over $\tau \equiv \tf - \ti$. The remaining contributions can easily be read off \fign{fig10}, where the upper leg is now denoted ``1'' while the lower leg is ``2'' (i.e. both have support in the amplitude).
The quantity we need is a 3-point function that lives during the time interval $[\ti+\tau,\ti]$. It is explicitly calculated in Eqs.~(\ref{eq:3-point-fin}) and (\ref{eq:3-point-fct}), and reads
\begin{align}
&\frac{\omega^2}{N_c^2-1} \big\langle  \rmtr  \left(V^\dag_2 \tmat^b V_1  \tmat^a\right) \Wc^{ba}(\x,\tf;\y,\ti) \rangle  \nn
&=\rme^{i\frac{\omega (\n_1^2 - \n_2^2)}{2} \tf } (\bdel_{x}- i \omega\n_2)\cdot (\bdel_{y}+i\omega \n_1) \tilde S^{(3)}(\x-\x_1(\tf),\y-\x_1(\ti),\v) \rme^{ i \omega \n_1 \cdot   (\x-\y- \Delta \x_{1})} \nn
&=\rme^{i\frac{\omega (\n_1^2 - \n_2^2)}{2} \tf  + i \omega \n_1 \cdot   (\x-\y- \Delta \x_{1})} (\bdel_{x} + i \omega\n_{12})\cdot \bdel_{y} \tilde S^{(3)}(\x-\x_1(\tf),\y-\x_1(\ti),\v) \,,
\end{align}
where $\Delta \x_1 = \x_1 (\tf) - \x_1(\ti)$ and $\v(s) = \n_{12}\, s$ describes the evolution of the size of the antenna in the relevant time interval.
In our case, the boundary conditions are $\y=\x_1(\ti)$ and $\x=\x_2(\tf)$, which leads to
\beq
\label{eq:sigma12}
 \Sigma_{12}(t) = \alpha_s C_F  \int_0^\infty  \rmd \tau   \int_0^ \infty \frac{\rmd \omega}{\omega^3}\, \rme^{ - i \frac{\omega}{2} \n_{12}^2 (t+\tau)} \bdel_{y}\cdot (\bdel_{x}+ i \omega\n_{12}) \tilde S^{(3)}(\x,\y,\v)\Big|_{\substack{\x=-\n_{12}(t+\tau) \\ \hspace{-3.5em} \y=\0}} \,.
\eeq
At this point we can make use of the time separation $\tau \ll t$ to simplify the above formula further. This was discussed extensively in Sec.~\ref{sec:interference}. We rewrite the phase in terms of the formation time of the gluon to see that
\be
\label{eq:small-angle-approximation}
\omega  \n_{12}^2 \ti \, \sim\,  \frac{ \x^2_{12}\,\tau}{ \r^2\, \ti} \, \ll \, \frac{\x^2_{12}}{ \r^2} \, \lesssim \, 1\,.
\ee
Therefore, unless the radiated gluon is collinear to either the quark or the gluon, the phase in \eqn{eq:sigma12} is negligible. Likewise, 
\beq
\bdel_{x}+ i \omega\n_{12} \approx  \bdel_{x} \,,
\eeq
based on the same parametric estimates.
Moreover, one can neglect the variation of $\v$ during the exchange, $\v(\ti) \approx \v(\tf) \equiv \x_{12}$. 

The remaining three contributions can be read off \fign{fig10}, where the upper leg is ``1'' and the lower leg is ``2''.
Hence, within the approximation described above, the sum over all four contributions reads
\beq
 \Sigma_{12}(t) = \alpha_s C_F  \int_0^\infty  \rmd \tau   \int_0^ \infty \frac{\rmd \omega}{\omega^3}\, K(\x_{12},\tau),
 \eeq
 where
\begin{align}
K(\x_{12},\tau)& = \bdel_x\cdot \bdel_y \tilde S^{(3)}(\x,\y, \x_{21}) \Big|_{\x=-\x_{21}, \y=\0} - \bdel_x\cdot \bdel_y \tilde S^{(3)}(\x,\y, \x_{21}) \Big|_{\x=\0, \y=\0} + (\x \leftrightarrow \y) \nn
&=  2\rmR \left\{\bdel_x\cdot \bdel_y \tilde S^{(3)}(\x,\y, \x_{12}) \Big|_{\x=-\x_{21}, \y=\0} - \bdel_x\cdot \bdel_y \tilde S^{(3)}(\x,\y, \x_{21}) \Big|_{\x=\0, \y=\0} \right\} \,.
\end{align}
where the three-point function $\tilde S^{(3)}(\x,\y,\x_{12})$ can be found in \eqn{eq:3-point-fct-ho-4}. Applying the Fourier transform, this corresponds to 
\beq
K(\x_{12},\tau) = 2 \rmR\,\int_{\k,\q,\l} \left(\rme^{i(\l -\k)\cdot \x_{12}} - \rme^{i \l \cdot \x_{12}} \right) (\k\cdot \q)\, \tilde S^{(3)}(\k,\q,\l) \,,
\eeq
where this time the relevant  3-point function can be read off  \eqn{eq:3-point-fct-ho-FT2}.
Expanding in the opening angle, this term can be written as 
\beq
 \Sigma_{12}(t) 
\simeq - \frac{1}{4} \Delta \hat q_{_F} (\x_{12},t)\,  \x_{12}^2,
\eeq
where we explicitly denote that the jet quenching parameter is proportional to the color factor of the fundamental representation, with 
\begin{align}
\label{eq:double-log-corrections1}
\Delta \hat q_{_F} (\x_{12},t) &=  \frac{\alpha_s C_F}{\pi} \, 2\rmR \int \dd \tau \int \dd \omega \, \frac{i \Omega^3}{\sinh \Omega \tau} \left[1+ \frac{4}{\sinh^2 \Omega \tau} \right] \nn
&= \frac{\alpha_s C_F}{\pi} \int_{\tau_0}^{\tau_\text{max}} \frac{\dd \tau}{\tau} \int^{\x^{-2}_{12}}_{\hat q \tau} \frac{\dd \q^2}{\q^2}\, \hat q \,.
\end{align}
Here we have stated the integration limits of the double logarithmic phase-space and regulated the double divergence. As expected, there is no vacuum contribution to this quantity.
In \eqn{eq:double-log-corrections1} $\tau_0 \sim m_{_D}^{-1}$ is related to the in-medium correlation length (where $m_D$ is the Debye screening mass) and $\tau_\text{max}$ is related to the time $t$. However, we have to make sure that the phase space for the integral over the transverse momentum is available, 
at any given time $t$ ($\tau_\text{max}$ is always limited by $L$ from above).
Since $t \leq \tdecoh$ (where $t_\text{d} \sim 1\big/ (\x^2_{12} \hat q)$ is the decoherence time), 
we can safely set $\tau_\text{max} = t$, and
\beq
\label{eq:double-log-corrections2}
\Delta \hat q_{_F} (\x_{12},t) = \frac{ \alpha_s C_F}{2 \pi}   \hat q_{_F} \ln \frac{t}{\tau_0} \left[ 2\ln\frac{1}{\x^2_{12}(t) m_{_D}^2} - \ln \frac{t}{\tau_0}\right] \,.
\eeq
The double-log contributions are the largest at $t=\tdecoh$, which leads to the constraint
\beq
\label{eq:double-log-limit}
\Delta \hat q_{_F} (\x_{12},t) \leq \frac{ \alpha_s C_F}{2\pi}   \hat q_{_F} \ln^2 \frac{1}{\x^2_{12}(t) m^2_{_D}} \,.
\eeq
In this suggestive form, it becomes clear that $\Sigma_{12}(t)$ accounts for a radiative correction to the dipole cross-section in \eqn{eq:dipole-xsection}, that can be absorbed into the jet quenching parameter $\hat q \to \hat q^{(1)} \equiv \hat q + \Delta \hat q(\x_{12},t)$. Those double logarithmic corrections in \eqn{eq:double-log-corrections2} are universal and have already been encountered in the context of transverse momentum broadening or radiative energy loss \cite{Liou:2013qya,Blaizot:2014bha,Wu:2014nca}.

We note that the radiative corrections which will accompany the (general) 3-point $\tilde S^{(3)}$, e.g. in \eqn{eq:spect-general-2} (and also for the interferences in \eqn{eq:antenna-spect}), that were calculated for the first time in \cite{Blaizot:2014bha}, can be implemented in our work by a similar redefinition of $\hat q$ in the respective spectra.

The full dipole cross-section reads then
\beq
S_1(\tend,0)  = \exp\left\{-\frac{1}{4} \int_0^\tend \rmd t\, \hat q_{_F}^{(1)}(\x_{12},t)\, \x^2_{12}(t) \right\} \,,
\eeq
such that the decoherence parameter is
\beq
\Delta_\sM (\tend) = 1- \big[ S_1(\tend,0) \big]^2 =1-  \exp\left\{ -\frac{1}{4} \int_0^\tend \rmd t\,\hat q_{_A}^{(1)}(\x_{12},t) \,\x^2_{12}(t) \right\} \,,
\eeq
where we used that $q_{_A} \approx 2 q_{_F} $ in the large-$N_c$ limit in order to compare to the leading-order decoherence parameter \cite{MehtarTani:2011tz}. 
This factor accounts for an {\sl accumulative} process of color decoherence as the pair propagates in the medium.

In summary, calculating the off-diagonal color structure $S_2$ at large-$N_c$, we have recovered the decoherence parameter describing an antenna traversing the medium and found how radiative corrections contribute. These novel ingredients adds to the program of computing radiative corrections to in-medium processes (transverse momentum broadening, medium-induced emissions and in-medium decoherence).

%%%%%%%%%%%%%%%%%%%%%%%%%%%%%
\subsection{Final answer and generalization to arbitrary color representation}
\label{sec:final-answer}
%%%%%%%%%%%%%%%%%%%%%%%%%%%%%

Combining the results obtained in Secs.~\ref{sec:direct-emissions}, \ref{sec:interference} and \ref{sec:radiative-corrections}, we find that the two-prong energy loss probability obeys the equation 
\begin{align}
\ProbSing(\epsilon, \tend)&= \delta(\epsilon)+ \int_0^\tend \rmd t   \int_0^\infty  \rmd  \omega\,  \big[ \Gamma_{11}(\omega,t) + \Gamma_{22} (\omega,t) \big] \,   \ProbTwo(\epsilon-\omega,t)  \nn
&+  \int_0^\tend \rmd t\,\big[ 1- \Delta_\text{med}(t) \big]  \int_0^\infty \dd \omega  \big[ \Gamma_{12}(\omega,t) + \Gamma_{21}(\omega,t) \big] \,\delta(\epsilon - \omega) \,.
\end{align}
Taking the derivative with respect to $t$ we obtain
\begin{align}
\label{eq:P12-evolution-eq}
\frac{\del}{\del t} \ProbSing(\epsilon,t) &=  \int_0^\infty  \rmd  \omega\,  \sum_{i} \Gamma_{ii}(\omega,t)\,   \ProbSing(\epsilon-\omega,t) \nn
& + \int_0^\infty \dd \omega\,\big[ 1- \Delta_\text{med}(t) \big]  \sum_{i\neq j} \Gamma_{ij}(\omega,t) \delta(\epsilon - \omega) \,,
\end{align}
which is identical to the expression one obtains after applying a time derivative to Eq.~(\ref{eq:main-result}).

To complement the knowledge about the quenching weight in Laplace space, we apply the same transformation to the two-pronged energy loss distributions, where the evolution equation (\ref{eq:P12-evolution-eq}) takes the following form
\beq
\label{eq:P12-evolution-eq-Mellin}
\frac{\del}{\del t} \ProbSingM(\nu,t)  = \gamma_\text{dir}(\nu,t) \ProbSingM(\nu,t) + \gamma_\text{int}(\nu,t) \,,
\eeq
where
\begin{align}
\gamma_\text{dir} (\nu,t) &=   \int_0^\infty \dd \omega \, \big(\Gamma_{11}(\omega,t) + \Gamma_{22}(\omega,t) \big) \rme^{-\nu \omega} \,, \\
\gamma_\text{int} (\nu,t) &=  \big[ 1- \Delta_\text{med}(t) \big]  \int_0^\infty \dd \omega \, \big(\Gamma_{12}(\omega,t) + \Gamma_{21}(\omega,t) \big) \rme^{-\nu \omega}  \,,
\end{align}
are the Laplace transformed rates for direct and interference radiation. Equation~(\ref{eq:P12-evolution-eq-Mellin}) is a nonhomogeneous, linear differential equation with initial condition $\ProbTwoM(\nu,0) = 1$, and is easily solved by 
\beq
\label{eq:two-prong-mellin-solution}
\ProbSingM(\nu,t)  =  \ProbOneM^2 (\nu,t) \left( 1+ \int_0^t \dd t' \, \gamma_\text{int}(\nu,t') \, \ProbOneM^{-2}(\nu,t') \right) \,,
\eeq
where the one-pronged energy loss distribution is given in \eqn{eq:ProbOneMellin}.

Explicitly, the direct rate is simply twice the rate for a single-prong,
\beq
\label{eq:direct-rate}
\gamma_\text{dir}(\nu,t) =4 \sqrt{\bar \alpha^2 \pi \nu \hat q} \,.
\eeq
The interference rate is also a sum of two equal terms, due to the equivalence of exchanging legs $1 \leftrightarrow 2$. We find that
\beq
\label{eq:interference-rate}
\gamma_\text{int}(\nu,t) &\simeq -2 \big[1-\Delta_\text{med}(t) \big] \gamma_\text{dir}(\nu,t) \,,
\eeq
for the relevant regime of large angle emissions, $\theta_\text{f} \gg \theta_{12}$.

%%%%%%%%%%%%%%%%%%%%%%%%%%%%%
\begin{figure}[t!]
\begin{center}
\includegraphics[width=0.5\textwidth]{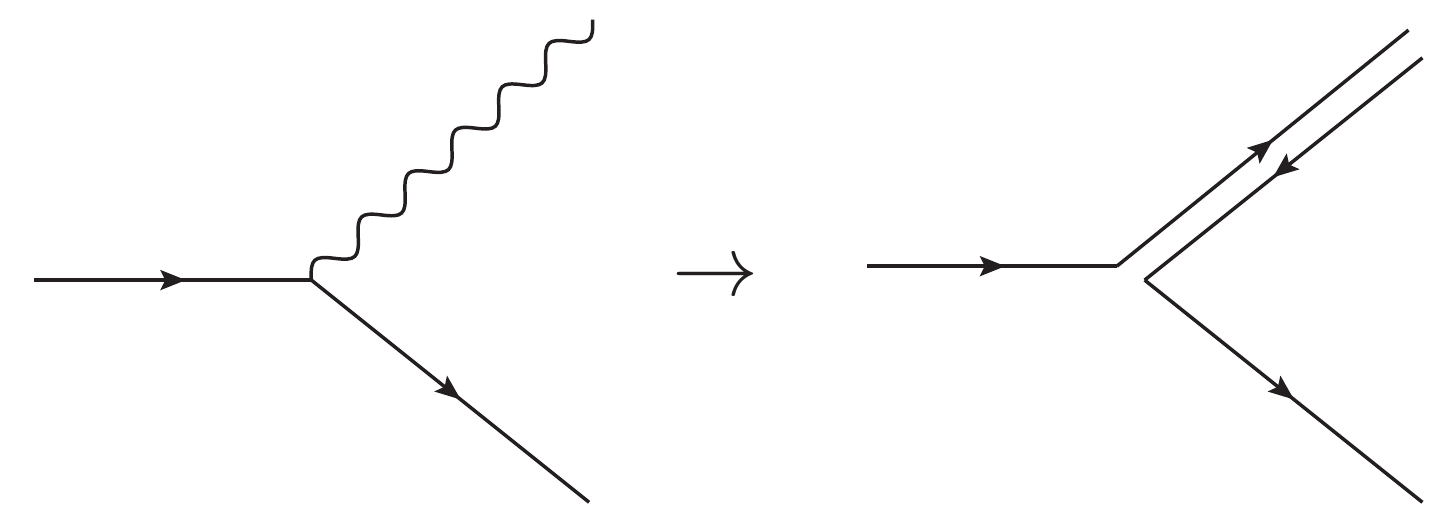}\\
\vspace{0.5cm}
\includegraphics[width=0.5\textwidth]{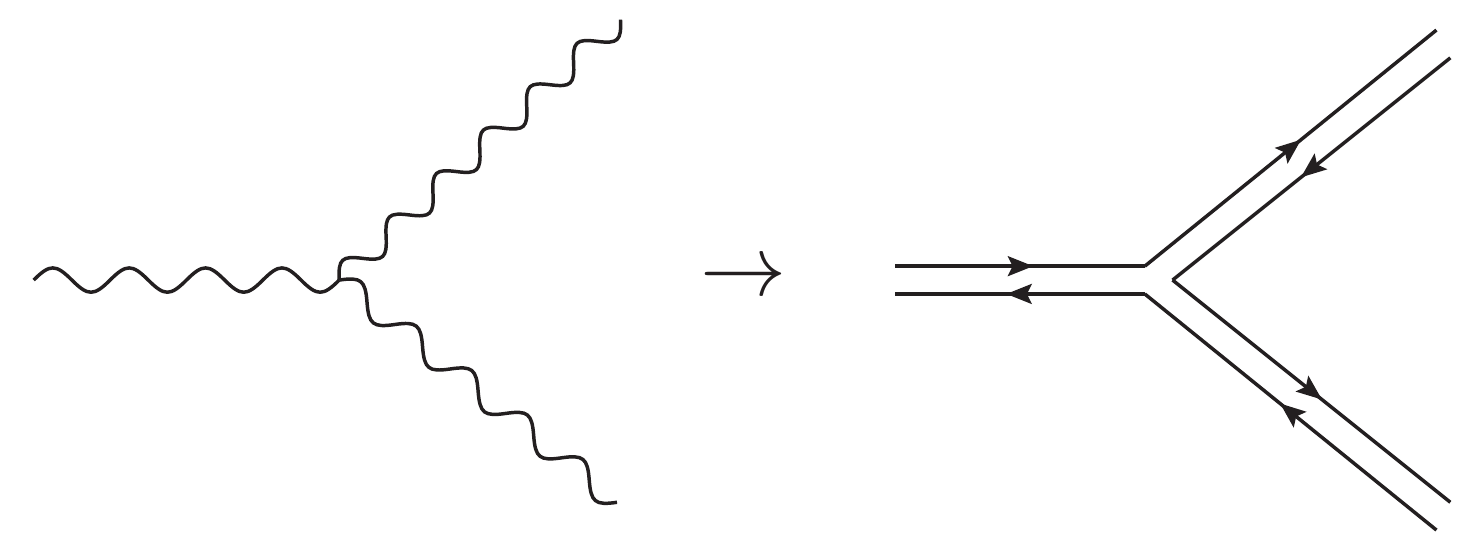}% 
\caption{Quark (upper) and gluon (lower diagram) splitting in the double-line (large-$N_c$) notation.}
\label{fig:splitting}
\end{center}
\end{figure}
%%%%%%%%%%%%%%%%%%%%%%%%%%%%%

\begin{figure}[t!]
\begin{center}
\includegraphics[width=0.45\textwidth]{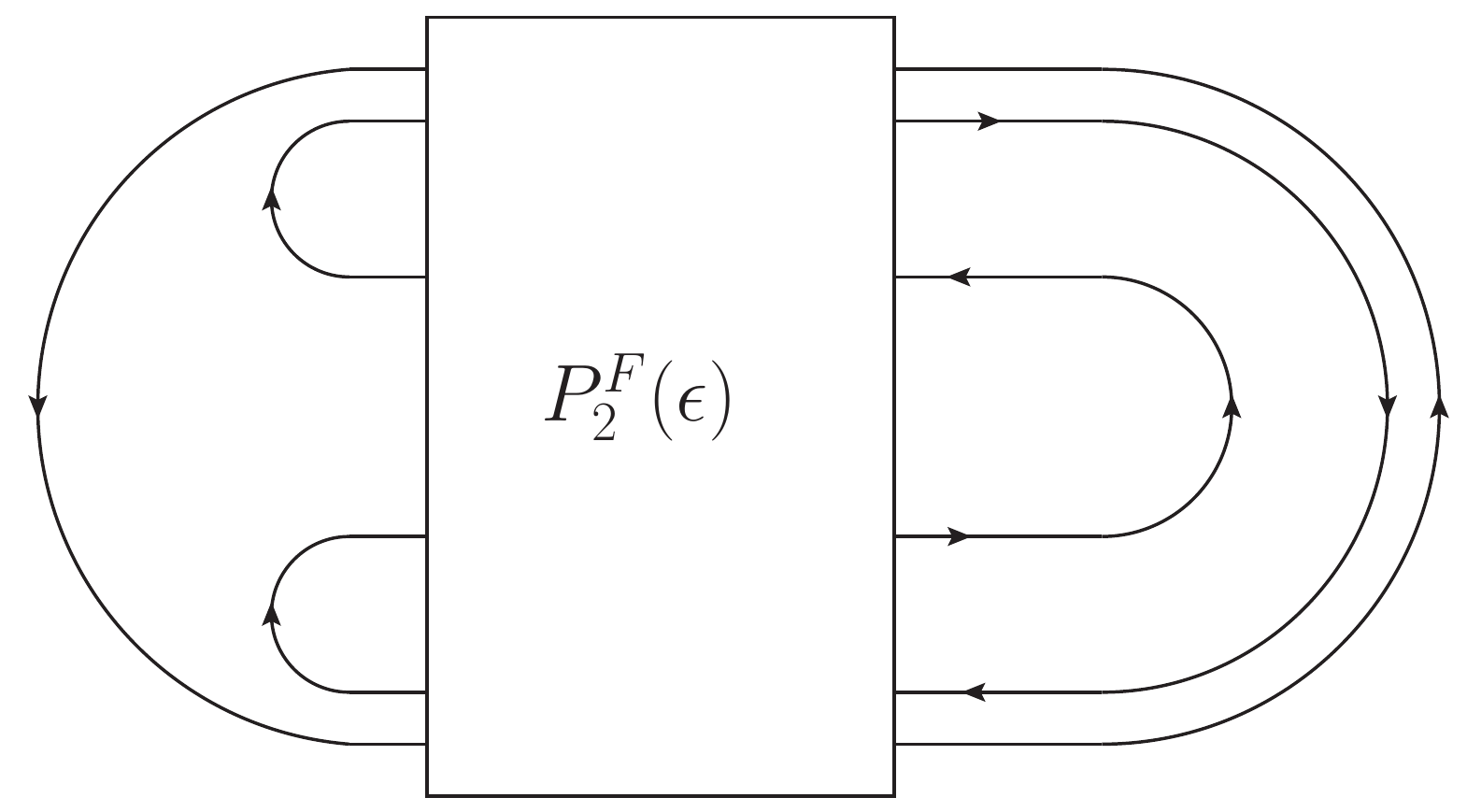}\qquad
\includegraphics[width=0.45\textwidth]{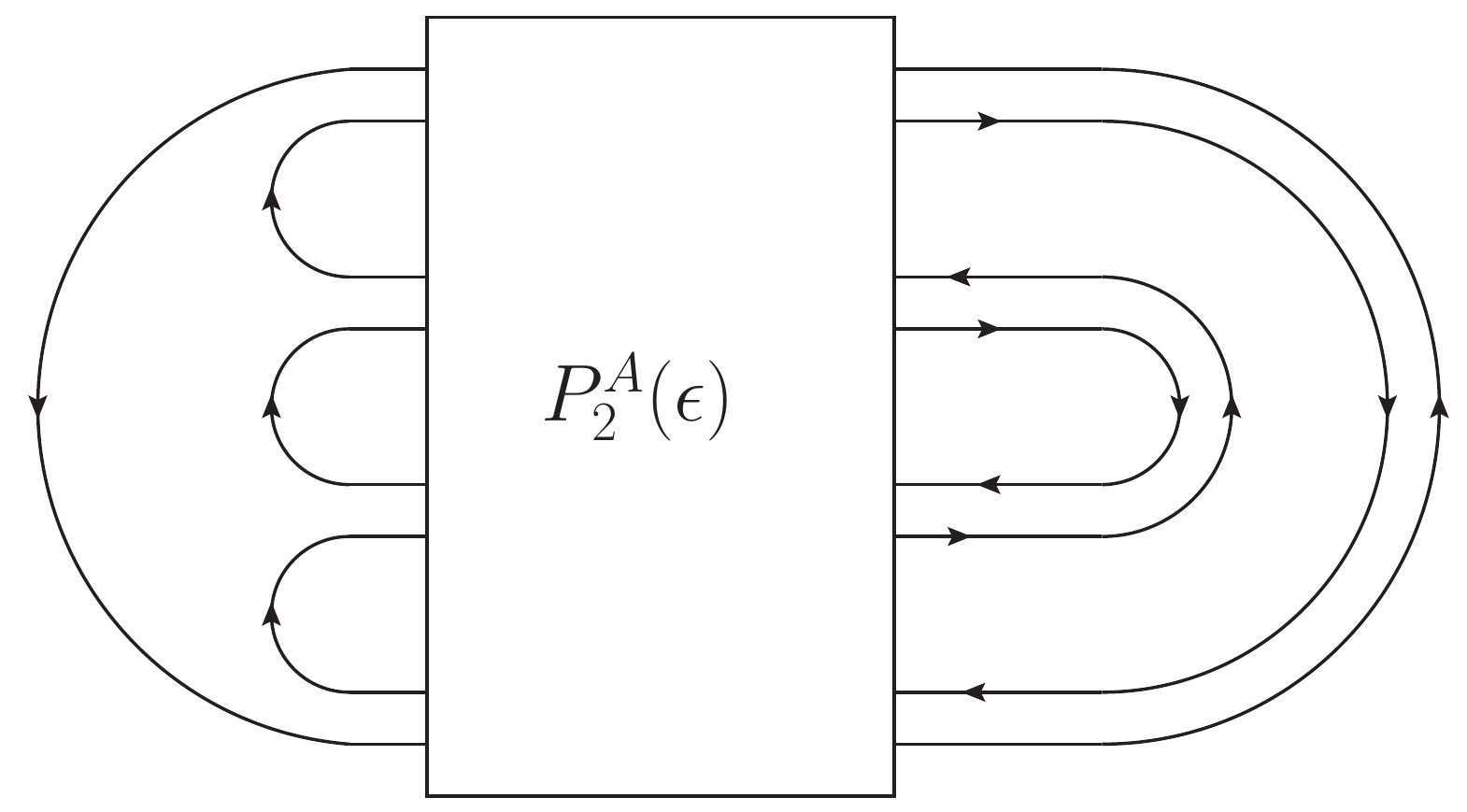}% 
\caption{The two-pronged energy loss distribution in fundamental (left) and adjoint (right diagram) color representation.}
\label{fig:quark-gluon-two-prong-energy-loss}
\end{center}
\end{figure}

Let us now consider the generalization of \eqn{eq:main-result} to arbitrary color representation of the initial projectile. From the outset we will neglect $g \to q+ \bar q$ splitting for two main reasons. First, because in this case all interferences between the outgoing quarks are suppressed in the large-$N_c$ limit and, second, because this splitting does not involve a soft divergence and will not contribute at leading-double-log accuracy in vacuum emissions. Hence, we are left with the processes $q \to g+ q$ and $g \to g+g$, both involving gluon emission off a total charge. Both also have soft and collinear divergences. We have depicted these fundamental splitting processes in \fign{fig:splitting} as Feynman diagrams and in the double-line notation that embodies the large-$N_c$ limit. The resulting diagrams for the two-pronged energy loss probability in the fundamental (quark splitting) and adjoint (gluon splitting) representations are depicted in \fign{fig:quark-gluon-two-prong-energy-loss}.

%%%%%%%%%%%%%%%%%%%%%%%%%%%%%
%%%%%%%%%%%%%%%%%%%%%%%%%%%%%

The generalization is straightforward in the large-$N_c$ limit. Both diagrams in \fign{fig:quark-gluon-two-prong-energy-loss} consists of two (for the fundamental) or three (for the adjoint) separately propagating color structures. Interferences can only arise from the central structure common to both and also to the color singlet, see \fign{fig8}, that has been extensively analysed above. Hence, additional diagrams arise only as direct terms involving the outer-most (for the fundamental antenna) and outer-most and inner-most (for the adjoint antenna) structures that only contribute to the direct emissions. Hence, in contrast to the color singlet case, a colored antenna can emit via the total color charge up to the flip.

The two-pronged energy loss probability for an antenna in arbitrary color representation will then consist of the following elements. In the completely incoherent case, the antenna can radiate off the total color charge or off the additional hard gluon. In term containing the interference contribution, the antenna behaves as the total charge right up to the flip. Hence, the two-pronged energy loss for arbitrary color representation reads,
\begin{align}
\label{eq:main-result-colored-1}
\ProbTwo^{\sst R}(\epsilon, L) 
&= \int_0^\infty\rmd \epsilon_1 \int_0^\infty \rmd \epsilon_2 \,  \ProbOne^{\sst R}(\epsilon_1, L) \, \ProbOne^{\sst A}(\epsilon_2, L)\, \delta(\epsilon-\epsilon_1-\epsilon_2) \nn 
& + \int_0^L \rmd t  \int_0^\infty\rmd \epsilon_0 \int_0^\infty\rmd \epsilon_1 \int_0^\infty \rmd \epsilon_2 \, \ProbOne^{\sst R}(\epsilon_1, L-t) \, \ProbOne^{\sst A}(\epsilon_2, L-t) \ProbOne^{\sst R}(\epsilon_0,t) \nn
& \times  \Big[ 1- \Delta_\sM(t) \Big] \int_0^\infty \rmd \omega  \,\sum_{i \neq j}\Gamma_{ij}(\omega, t) \, \delta(\epsilon-\epsilon_0- \epsilon_1-\epsilon_2-\omega) \,,
\end{align}
where $R = F/A$ and we define $\ProbOne^{\sst R}(\epsilon,t)$ according to \eqn{eq:one-prong-eloss} with the substitution $\omega_s \to \omega_s^{\sst R} \equiv \frac{C_{\sst R}^2}{C_F^2} \omega_s$. Finally, the color factor of the interference spectrum, see \eqn{eq:bdmps-interference}, entering the rates $\Gamma_{12}(\omega,t)$ and $\Gamma_{21}(\omega,t)$ has to be transformed as $-C_F \to -N_c/2$.\footnote{This is a consequence of color conservation. Consider a total charge $\Q_0$ splitting into two daughter partons with charges $\Q_1$ and $\Q_2$, respectively, so that $\Q_0 = \Q_1 + \Q_2$. Interferences will be proportional to the product of daughter charges, which are found to be $\Q_1\cdot \Q_2 = (\Q_0^2 - \Q_1^2 - \Q_2^2)/2$, where $\Q^2 \equiv C_{\sst R}$. For photon splitting $\Q_1\cdot \Q_2 = -C_F$, while for gluon emission $\Q_1\cdot \Q_2 = - N_c/2$.} This more general distribution, follows from the evolution equation
\beq
\label{eq:P12-evolution-eq-Mellin-colored}
\frac{\del}{\del t} \ProbTwoM^{\sst R}(\nu,t)  = \gamma_\text{dir}(\nu,t) \ProbTwoM^{\sst R}(\nu,t) + \gamma_\text{int}(\nu,t) \ProbOneM^{\sst R}(\nu,t) \,,
\eeq
which is solved by
\beq
\label{eq:two-prong-mellin-solution-colored}
\ProbTwoM^{\sst R}(\nu,t)  =  \ProbOneM^{\sst R} (\nu,t) \ProbOneM^{\sst A}(\nu,t) \left( 1+ \int_0^t \dd t' \, \gamma_\text{int}(\nu,t') \ProbOneM^{-1,\sst A}(\nu,t') \right) \,,
\eeq
where the initial condition was set as $\ProbTwoM^{\sst R}(\nu,0) = 1$. Applying that $C_F \simeq N_c/2$, consistent with the large-$N_c$ approximation, we find that $\ProbTwoM^{\sst R}(\nu,t) = \ProbOneM^{\sst R} (\nu,t) \ProbSingM (\nu,t)$, which leads to \eqn{eq:main-result-colored}.

%%%%%%%%%%%%%%%%%%%%%%%%%%%%%
\section{Conclusions and outlook}
\label{sec:conclusions}
%%%%%%%%%%%%%%%%%%%%%%%%%%%%%

In this work, we have computed the energy loss probability distribution of a boosted pair of partons created promptly in a dense  medium. Our main results are given in \eqn{eq:main-result} and \eqn{eq:main-result-colored} and are derived in the large-$N_c$ approximation. We have identified two processes that delay the independent energy degradation of the individual partons, related to quantum and color decoherence. In effect, the delay is related to the time when the color of the dipole ``flips'', see Secs.~\ref{sec:interference} and \ref{sec:radiative-corrections}, induced by an interference exchange. 
Before the flip, then, the medium only resolves the total color charge of the system and the dipole loses energy {\it as if it were the parent parton}. After the flip, on the other hand, energy is lost independently by the dipole constituents.

The two-pronged energy loss distribution is a new tool that can be applied to observables involving vacuum splittings which take place inside the medium, albeit within the approximation that the formation time is much shorter than the decoherence time. 
It is important for phenomenological applications of jet quenching to capture these effects due to the collinear singularity of vacuum emissions inside a jet. Hence the theoretical uncertainties of jet quenching observables that are affected by such splittings can be greatly reduced. Simply considering the small- and large-angle limits of our expressions, immediately imply a {\it collimation} of the jet sample emerging in heavy-ion collisions compared to proton-proton due to the additional suppression of large-angle structures. Our approach may be extended to multiple vacuum splitting to compute radiative energy loss of the fluctuating jet substructure.  Phenomenological applications of this tool will be presented in upcoming publications.

As we already have pointed out, the vacuum splitting process at the cause of the two-pronged structure is non-local from the point of view of medium interactions (decoherence). However, it is amenable to a probabilistic interpretation where the angular and time-scales are clearly defined from the kinematics of the parent dipole. Our results will therefore also serve as guidance for current and future Monte-Carlo event generators of jet quenching.

%%%%%%%%%%%%%%%%%%%%%%%%%%%%%
\section*{Acknowledgements}
%%%%%%%%%%%%%%%%%%%%%%%%%%%%%

We would like to thank Jean-Paul Blaizot, Edmond Iancu, Guilherme Milhano,  Al Mueller, Carlos A. Salgado, Soeren Schlichting and Gregory Soyez  for interesting discussions. 
KT has been supported by a Marie Sklodowska-Curie Individual Fellowship of the European Commission's Horizon 2020 Programme under contract number 655279 ResolvedJetsHIC. The research of YMT is supported by the U.S. Department of Energy under Contract No. DE-FG02-00ER41132.

\appendix

%%%%%%%%%%%%%%%%%%%%%%%%%%%%%
\section{Feynman rules for LCPT in a background field}
\label{sec:FeynmanRules}
%%%%%%%%%%%%%%%%%%%%%%%%%%%%%

We effectively work in light-cone perturbation theory (LCPT) where the time-integrations are kept explicit. In vacuum, integrating out time would lead to the well-known procedures of energy denominators. In the presence of a background field, keeping time-ordering explicit allows to resum multiple interactions.

Let us summarize the corresponding Feynman rules. Each internal propagator must be multiplied with a factor $1/(2 E)$ where $E$ is the conserved $+$-component of the momentum along the eikonal trajectory. In deriving the propagators in the mixed representation , that is, the 3-momentum $(p^+,\p)\equiv (E,\p)$ and time $x^+\equiv t$ (see Appendix A in Ref.~\cite{Blaizot:2012fh}), we have neglected the instantaneous part that is only present in the free part of the propagator. 
Hence, for each internal quark (antiquark) line we have
\beq
\frac{1}{2E}\,\Gc_{\sst F (\bar F)}^{ji}(\p_2,t;\p_1,t_0 | E)\, \delta^{ss'} \,,
\eeq
which has transverse momentum $\p_1$ ($\p_2$) and initial (final) time $t$ ($t_0$) and 
where $i$ and $j$, are the initial and final quark colors and, $s$ and $s'$ their respective helicities.
For a gluon, 
\beq
\frac{1}{2E}\,\Gc_{\sst A}^{ba}(\p_2,t;\p_1,t_0|E) \,\delta^{\lambda\lambda'} \,,
\eeq
where $\lambda,\lambda'$ refer to the initial and final polarizations. External propagators do not have the factor $1/(2E)$ and have to be multiplied by an additional phase, which falls out when taking the square of the amplitude and can therefore be neglected. Suppressing the explicit color indices, the expression for the propagator is given explicitly in \eqn{eq:greens-function} for arbitrary color representation.

In the limit  $ 2E/\p^2 \gg t\sim L $, known as the shock wave limit the kinetic term can be neglected and the propagator reduces a Wilson line whose trajectory is frozen in transverse space,
\beq\labe{eq:prop-eiko}
\Gc(\x,t; \x_0,t_0) \,  \to \, \delta(\x-\x_0)\, U(t,t_0;[\x]) \,.
\eeq
Keeping instead the information about the finite ``tilt'' of the Wilson line, we obtain 
\beq
\Gc(\x,t; \x_0,t_0) = \Gc_0(\x,t; \x_0,t_0)  U(t,t_0;[\x_\text{cl}]) \,,
\eeq
where the classical trajectory is given by $\x_\text{cl}(s) = \x_0 + \frac{s-t_0}{t-t_0}(\x_1 - \x_0)$ and $\Gc_0$ is the vacuum propagator in coordinate space,
\beq
 \Gc_0(\x,t; \x_0,t_0) = \Theta(t-t_0)\frac{E}{2\pi i \,(t-t_0)} \exp\left[i\frac{E}{2}\frac{(\x-\x_0)^2}{t-t_0} \right] \,.
\eeq
After performing the Fourier transforms with respect to the end-points of the propagator, we obtain
\beq
\Gc(\p,t; \p_0,t_0) = \rme^{-i \frac{\p^2}{2E}(t-t_0)} \int_{\y_0,\y_1}  \rme^{-i(\p-\p_0)\cdot (\y_0 + t_0 \y_1)} \eta_\varepsilon\left( \y_1 - \frac{\p}{E} \right) U\Big( \big[\x_\text{cl}(s) = \y_0 + s \y_1 \big] \Big) \,,
\eeq
where $\eta_\varepsilon(\x) = \rme^{\x^2/(2\varepsilon)}/(2\pi \varepsilon)$ is the heat kernel in two dimensions and embodies the Heisenberg uncertainty principle. In our case $\varepsilon \equiv i/(E(t-t_0))$. It is also a nascent delta function, $\delta(\x) = \lim_{\varepsilon \to 0} \eta_\varepsilon(\x)$, so taking the limits $E \gg (t-t_0)^{-1}$ and $(\p/E) s \gg \y_0 $, we obtain
\beq
\label{eq:TiltedWilsonLine}
\Gc(\p,t; \p_0,t_0) = (2\pi)^2 \delta(\p-\p_0)\, U\big(t,t_0;[\x_\text{cl}(s) = \n s  ] \big) \, \rme^{-i \frac{\p^2}{2E}(t-t_0)},
\eeq
where the two-dimensional vector $\n=\p/E$ parameterizes the trajectory of the projectile.

%%%%%%%%%%%%%%%%%%%%%%%%%%%%%
\begin{figure}[t!]
\begin{center}
\includegraphics[width=12cm]{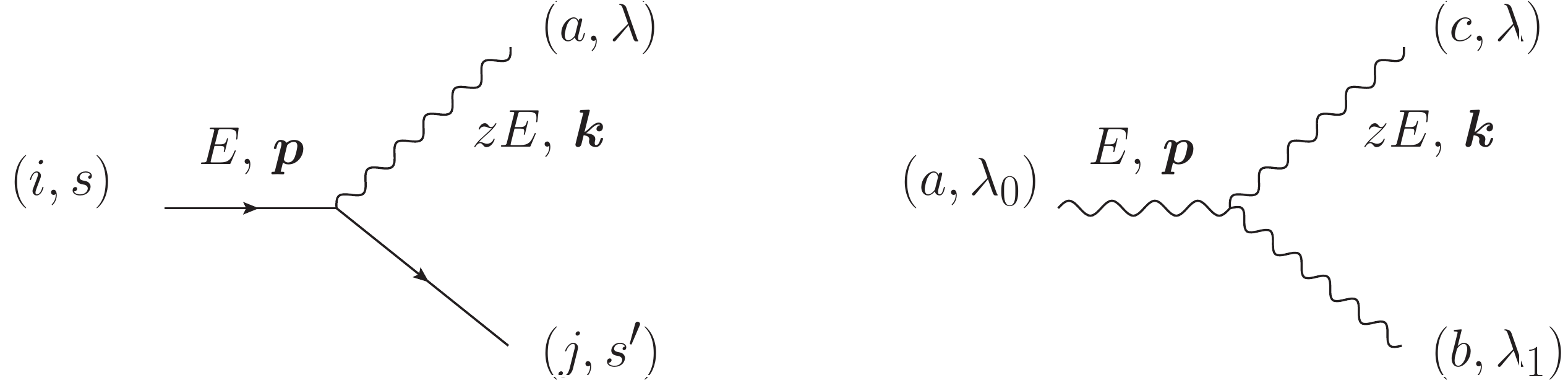} 
\caption{Quark and gluon splitting vertices. See text for details.  }
\label{fig:quark-split}
\end{center}
\end{figure}
%%%%%%%%%%%%%%%%%%%%%%%%%%%%%

Below, we list the splitting vertices necessary for the computation as depicted in Fig.~\ref{fig:quark-split}. Initial and final momenta are incoming and outgoing, respectively. The quark--gluon vertex reads,
\begin{align}
\labe{eq:q-g-vertex}
V_{ss'\lambda}(\q,z)&=(ig \,\tmat^a) \, \bar u(p',s')  \slashed{\epsilon}_\lambda^\ast(k) u(p,s) \nn
& = - \frac{2 i g\, \tmat^a}{z \sqrt{1-z}} \delta_{s's} \left[\delta_{ \lambda, s } + (1-z)\delta_{\lambda, -s } \right]\, \q\cdot\beps_\lambda^\ast  \,.
\end{align}
where $\q\equiv \k-z \p$. The antiquark--gluon vertex is obtain form the former by the replacement $i \to -i$ and $s \to -s$.
The gluon--gluon vertex reads
\begin{align}
\labe{eq:g-g-vertex}
&V_{\lambda_0\lambda_1\lambda}^{abc}(\q,z)=\nn
& - 2 i g \,(\Tmat^c)^{ab} \Bigg[\frac{1}{z}(\q\cdot\beps_\lambda^\ast) \,\delta_{\lambda_0 \lambda_1}  + \frac{1}{1-z}(\q\cdot\beps_{\lambda_1}^\ast)  \,\delta_{\lambda_0\lambda}  - (\q\cdot\beps_{\lambda_0})  \, \delta_{\lambda_1 \lambda}\Bigg]  \,.
\end{align}
Furthermore, we have the gluon--quark vertex
\begin{align}
\labe{eq:g-qq-vertex}
V_{\lambda ss'}(\q,z) &= (ig \,\tmat^a) \, \bar u(p',s')  \slashed{\epsilon}_\lambda (k) v(p,s) \nn
&= - \frac{2 i g \,\tmat^a}{ \sqrt{z(1-z)}} \delta_{-s's} \left[z \delta_{ \lambda, s } - (1-z)\delta_{\lambda, s' } \right]\, \q\cdot\beps_\lambda   \,.
\end{align}
Similarly, the  photon--quark is simply, 
\beq
\label{eq:photon-vertex}
A_{\lambda_0,s,s'}(\q,z)   =    \frac{2 i e }{\sqrt{z(1-z)}} \delta_{-s' s} \left[z \delta_{ \lambda, s } - (1-z)\delta_{\lambda, s' } \right]\, \q\cdot\beps_\lambda.
\eeq
Finally, to each vertex one must attach a time integration.

%%%%%%%%%%%%%%%%%%%%%%%%%%%%%
\section{Connecting to the hard vertex}
\label{sec:HardVertex}
%%%%%%%%%%%%%%%%%%%%%%%%%%%%%

In order to put the Feynman diagrams listed previously to use, we also have to define how to glue the propagators to the amplitude of the hard vertex. Since the process that creates the particle we are interested in is supposed to be very hard, the related time-scale is vanishingly small. In the simplest case, e.g., for the production of a single quark propagating in the medium, we simply can fix the initial time to $t_0 = 0$ and factorize the amplitude for the hard vertex (which only depends the energy and transverse momentum of the initial particle).

Consider as a concrete example the emission of a gluon, with energy $\omega \equiv zE$ and transverse momentum $\k$, off a quark, with correspondingly $\{E,\p \}$, in the medium. 
Using the Feynman rules listed in \ref{sec:FeynmanRules}, the amplitude reads
\begin{align}
\labe{eq:amp-mixed}
&\Mc^{(a,i)}_{(\lambda,s)}(p,k) = 
\int_{\k',\p',\p_0} \int_{0}^\infty \rmd t \, \Gc_{_A}^{ab} (\k,L; \k', t| zE)\frac{1}{2E} \nn 
& \times  \left[  \Gc_{_F}(\p,L; \p'-\k', t|(1-z)E)   \, V_{\lambda,s,s'}^b(\k' - z \p',z) \,  \Gc_{_F} (\p',t; \p_0, 0|E)   \right]^{ij} \mathcal{M}^j_{s'}(p_0)\,.
\end{align}
By performing the soft gluon approximation, $z \ll 1$, in the vertex and neglect the transverse momentum broadening of the quark, see \eqn{eq:TiltedWilsonLine}, we get
\begin{align}
\labe{eq:amp-mixed-2}
\Mc^{(a,i)}_{(\lambda,s)}(p,k) &=-\frac{i g}{\omega} \int_{\k'} \int_0^L \rmd t \, \, \rme^{-i \n\cdot \k' t +i\frac{\omega  }{2 }\n^2 t}\,\,  (\k' -\omega \n ) \cdot\beps^\ast_\lambda\nn
& \times   \Gc_{_A}^{ab} (\k,L; \k', t|\omega)  \left[  V(L,t)    \tmat^b\,  V(t, 0)  \right]^{ij}   \Mc^{j}_{s}(p) \,,
\end{align}
up to (irrelevant) phase factors, 
where we have approximated $\p' \approx \p$ and $V \equiv U_{_F}(\x)$ stands for a Wilson-line in the fundamental representation evaluated along the classical trajectory of the hard emitting parton and $\n\equiv \p/E$.
We also note that the  $\k'$ integration is equivalent to Fourier transforming the propagator, namely, 
\beq
 \int_{\k'} \rme^{-i \n\cdot \k't }\,  (\k' -\omega \n) \cdot\beps_\lambda^\ast \,  \, \, \Gc_{_A}^{ab} (\k,L; \k', t|\omega)= (i \bdel_{x}-\omega\n) \cdot \beps_\lambda^\ast \,   \left. \Gc_{_A}^{ab} (\k,L; \x, t|\omega) \right|_{\x = \n t } \,,
\eeq
which leads to the expression in \eqn{eq:amp-mixed-3}.

More insight is gained about the underlying vacuum process when considering what happens for an early creation of a time-like dipole (antenna), focussing initially on the color singlet case. For simplicity, we ignore longitudinally polarized photons that contribute to the instantaneous pair production and consider only transverse polarization.
Let us only write a part of the process, stopping at some time $t$ after the antenna has formed. The partial amplitude reads then
\begin{align}
M^{ij}_{0}(12) &= \int_{\p_1',\p'_2} \int_{0}^\infty \dd t_0 \left[ \Gc_F(\p_1,t; \p'_1,t_0 | zE) \bar \Gc_F(\p_2,t; \p'_2 ,t_0 | (1-z)E)  \right]^{ij} \nn
&\times  A_{\lambda_0,s_1,s_2}((1-z)\p'_1 - z \p'_2,z) \,\frac{1}{2E} \rme^{-i \frac{(\p'_1 + \p'_2)^2}{2 E}t_0} \mathcal{M}_{\lambda_0}(p'_1+p'_2) \,,
\end{align}
where the $\gamma^\ast \to q\bar q$ vertex is denoted by $A$, see \eqn{eq:photon-vertex}. We manipulate this expression slightly in order to reconstruct the right kinematics of this splitting, as
\begin{align}
M^{ij}_0(12) &=\int_{\p_1',\p'_2} \int_{0}^\infty \dd t_0 \left[ \Gc_F(\p_1,t; \p'_1,t_0 | zE) \bar \Gc_F(\p_2,t; \p'_2 ,t_0 | (1-z)E)  \right]^{ij}  \rme^{-i \frac{\p'^2_1}{2 zE} t_0 - i \frac{\p'^2_2}{2(1-z)E}t_0 } \nn
&\times  A_{\lambda_0,s_1,s_2}((1-z)\p'_1 - z \p'_2,z) \,\frac{1}{2E} \rme^{i \frac{[(1-z)\p'_1 - z\p'_2]^2}{2z(1-z) E}t_0} \mathcal{M}_{\lambda_0}(p'_1+p'_2) \,,
\end{align}
where $t_\gamma = 2z(1-z) E/[(1-z)\p'_1 - z\p'_2]^2$ is the photon formation time.
The  additional phase-factors associated with the two propagators limits the dependence on the initial time to determining the initial position of where the $q\bar q$ pair is formed, for the eikonal case see Eq.~(\ref{eq:TiltedWilsonLine}). Currently, we assume that the photon decays rapidly (for $E \to \infty$ and $z \sim 1/2$), so that the extent of $t_0$ is again severely limited and we can simply write
\beq
M^{ij}_0(12) = \left[ \Gc_F(\p_1,t; \p'_1,0 | zE) \bar \Gc_F(\p_2,t; \p'_2 ,0 | (1-z)E)  \right]^{ij} \mathcal{M}_{s_1,s_2} (p'_1,p'_2) \,,
\eeq
in effect factorizing the propagation and production processes, with
\begin{align}
\mathcal{M}_{s_1,s_2} (p_1,p_2) &=\frac{1}{2E} \int_0^\infty \dd t_0 \, \rme^{i \frac{[(1-z)\p_1 - z \p_2]^2}{2 z(1-z)E}t_0} \, A_{\lambda_0,s_1,s_2}((1-z)\p_1 - z \p_2,z) \mathcal{M}_{\lambda_0}(p_1+p_2) \nn
&= \frac{i}{E_1 E_2 \, (\n_1 - \n_2)^2} A_{\lambda_0,s_1,s_2}((1-z)\p_1 - z \p_2,z) \mathcal{M}_{\lambda_0}(p_1+p_2) \,,
\end{align}
being the amplitude describing a $\gamma^\ast \to q \bar q$ splitting in vacuum.
In vacuum, squaring the above amplitude summing over spin, color and flavor of the final state, see \eqn{eq:amplitude-squaring}, yields
\beq
\frac{\rmd \sigma_\text{vac}}{\rmd \Omega_1\rmd \Omega_2} = \frac{2 e^2}{ E_1 E_2 (\n_1 - \n_2)^2} P_{\gamma q}(z) \frac{\rmd \sigma_\text{vac}}{\rmd \Omega_0} \,,
\eeq 
where $E_1 \equiv z E$ ($E_2 \equiv (1-z) E$) and $\n_1 \equiv \p_1/E_1$ ($\n_2\equiv \p_2/E_2$) is the energy and direction of the quark (antiquark). Considering the frame $\p_1=-\p_2 \equiv \p$, leads to
\beq
\label{eq:photon-vac-spectrum}
\frac{\rmd \sigma_\text{vac}}{\rmd z \rmd E \, \rmd \p^2} =  \frac{\alpha_\text{em}}{2\pi} \frac{P_{\gamma q}(z)  }{\p^2} \frac{\rmd \sigma_\text{vac}}{\rmd E} \,,
\eeq 
where $P_{\gamma q}(z) =n_f N_c [z^2+(1-z)^2] $. This expression can be straightforwardly generalized to QCD splittings by substituting for the coupling $\alpha_\text{em} \to \alpha_s$ and replacing the Altarelli-Parisi splitting function by the desired one.

%%%%%%%%%%%%%%%%%%%%%%%%%%%%%
\section{Medium averaged 2-point and 3-point functions}
\label{sec:n-point-fct}
%%%%%%%%%%%%%%%%%%%%%%%%%%%%%

We encountered two types of correlation functions to be averaged over the ensemble of background field configurations. The first quantity is the 2-point function which reads 
\beq
( X_f| S^{(2)} |X_i ) =\frac{1}{N_c^2-1}\langle  \rmTr \,U^\dag_1  \, \Gc(\z_f,\tf; \z_i,\ti |\omega) \rangle,
\eeq
where the initial and final points are $X_i\equiv (\z_i,\x_1(\ti ))$ and $X_f\equiv (\z_f,\x_1(\tf))$.
The second quantity is the 3-point function
\beq
( X_f| S^{(3)} |X_i ) =\frac{2}{N_c^2-1}\langle  \rmtr \left(V^\dag_2 t^a V_1 t^b \right)  \, \Gc^{ab}(\z_f,\tf; \z_i, \ti|\omega) \rangle
\eeq
where here the initial and final points read $X_i\equiv (\z_i,\x_1(\ti);\x_2(\ti))$ and $X_f\equiv (\z_f, \x_1(\tf);\x_2(\tf))$.
It can be verified that when $\x_1 =\x_2$, the 3-point functions reduces to the 2-point function. Hence, one only needs to compute the latter. 

In the path integral formulation the 3-point function reads
\begin{align}
\label{eq:threepoint-pathintegral}
( X_f| S^{(3)} |X_i ) &=\int_i^f \Dc \r  \, \exp\left\{ \frac{i\omega}{2} \int_{t_i}^{t_f} \rmd s\, \dot \r^2(s) \right. \nn
&  \left.      -\frac{1}{4}\int_{t_i}^{t_f} \rmd s \left[N_c n\sigma(\r-\x_1)+N_c n\sigma(\r-\x_2)+(2 C_F-N_c) n\sigma(\x_1-\x_2)\right]\right\} \,.
\end{align}
We make now the following change of variables,
\beq
\u \equiv \r -\x_1 \quad \text{and}\quad \v=\x_1-\x_2 \,,
\eeq
which allows to show, that the 3-point function $S^{(3)}$ obeys the Schr\"odinger equation
\begin{align}
&S^{(3)}(\z,\y, \v;\tau) =\Gc^{(0)}(\z-\y,\tau)  \\
&+ \int_0^\tau \rmd t \int \rmd \u  \, \Gc^{(0)} (\z-\u, \tau-t)\frac{N_c n }{2} \left[ \sigma(\u)+\sigma(\v-\u)+ \left(2 \frac{C_F}{N_c}-1 \right) \sigma(\v)\right] S^{(3)}(\u,\y, \v; t) \,, \nonumber
\end{align}
where $\tau = \tf - \ti$ and $n$ denotes the density of scattering centers in the medium. 
Applying this to Eq.~(\ref{eq:threepoint-pathintegral}), we find that, as a result, the quantum phase yields (recall that $\x_1(s)= \n_1s$ and $\x_2(s)= \n_2 s$),
\begin{align}
&\int_{\ti}^{\tf} \rmd s\, \dot \r^2(s) = \int_{\ti}^{\tf} \rmd s\, 
\left(\dot \u(s)+\n_1\right)^2 \nn
&= \int_{\ti}^{\tf} \rmd s\, 
\dot \u^2(s) + 2 \big(\z_f- \x_1(\tf))\cdot \n_1- 2\big(\z_i- \x_1(\ti) \big)\cdot \n_1+ \n_1^2(\tf- \ti) \,.
\end{align}
Then the 3-point function therefore reads 
\beq
\label{eq:3-point-fin}
&&( X_f| S^{(3)} |X_i ) = \tilde S^{(3)}(\u_f,\u_i,\v) \exp\left\{ i\omega \frac{\n_1^2}{2} (\tf-\ti)+ i \omega \n_1 \cdot   (\u_f-\u_i) \right\},
\eeq
with 
\begin{align} 
\label{eq:3-point-fct}
& \tilde S^{(3)}(\u_f,\u_i,\v)  \equiv \nn
&\int  \Dc \u  \, \exp\left\{ \frac{i\omega}{2} \int_{\ti}^{\tf} \rmd s \, \dot\u^2 - \frac{N_c n}{4}\int_{\ti}^{\tf} \rmd s\, \left[\sigma(\u)+\sigma(\u+\v)+\left(2 \frac{C_{_F}}{N_c}-1
\right) \sigma(\v)\right]\right\} \,,
\end{align}
and $\u_f \equiv \z_f - \x_1(\tf)$, $\u_i \equiv \z_i - \x_1(\ti)$ and 
\beq\label{eq:v-def}
\v(t)= \n_{12} \, t,
\eeq 
where $\n_{12} = \n_1 - \n_2$, is the coordinate of the emitting system centre-of-mass. 
When $\v$ is constant (which we will assume when deriving the double logarithmic contribution) $\tilde S^{(3)}(\u_f,\u_i,\v)$ is equivalent to that introduced in Ref.~\cite{Blaizot:2014bha}. 
The 2-point function is deduced from the 3-point function by letting $\p_1=\p_2$:
\beq
\label{eq:2-point-fct}
( X_f| S^{(2)} |X_i ) = \tilde S^{(3)}(\u_f,\u_i,\0)  \exp\left\{ i \omega \frac{\n_1^2}{2} (\tf-\ti)+ i \omega \n_1 \cdot   (\u_f-\u_i) \right\}  \,. 
\eeq
To make contact with previous notations, we point out that $\tilde S^{(3)}(\u_f,\u_i,\0) \equiv \Kc (\u_f,\u_i)$, where $\mathcal{K}(\u_f,\u_i)$ is defined in \eqn{eq:ho-int}.

%%%%%%%%%%%%%%%%%%%%%%%%%%%%%
\section{The harmonic approximation }
\label{sec:harmonic-approx}
%%%%%%%%%%%%%%%%%%%%%%%%%%%%%

The reduced 3-point function \eqn{eq:3-point-fct} is the basic building block to be evaluated. This can be carried out analytically in the harmonic approximation,
\beq
\label{eq:harmonic-approx}
N_c n\,\sigma(\x) \simeq \frac{1}{2}\,\hat q_{_A}  \,\x^2 \,,
\eeq
where we have explicitly denoted the color factor dependence of the jet quenching parameter (throughout the paper $\hat q \equiv \hat q_{_A}$, unless explicitly stated otherwise).
Using \eqn{eq:harmonic-approx} in \eqn{eq:3-point-fct} and assuming $\v\approx \v(\tf) \approx \v(\ti)$ to be constant in the interval $\tau\equiv  t_f-t_i$, which is assumed to be small throughout the paper, we find
\beq
\label{eq:3-point-fct-ho-1}
\tilde S^{(3)}(\u_f,\u_i,\v)  &\simeq \exp\left[-\frac{1}{4}\left(\frac{C_{_R}}{N_c}-\frac{1}{4}\right) \hat q \, \v^2\,\tau\right] \mathcal{K}(\x_f,\x_i)
\eeq
where 
\beq\label{eq:x-def}
\x\equiv \u+\frac{\v}{2} \,.
\eeq
The resulting quadratic path integral in $\mathcal{K}(\x_f,\x_i)$ is standard and yields
\begin{align}
 \label{eq:ho-int}
 \mathcal{K}(\x_f,\x_i) &=\int \Dc \x  \, \exp\left[ \frac{i\omega}{2} \int_{\ti}^{\tf} \rmd s \, \left( \dot\x^2 + i \frac{\x^2}{2 \tform^2}\right) \right] \nn
&  = \frac{\omega \Omega}{2\pi i \, \sinh \Omega \tau} \, \exp\left\{ \frac{i\omega \Omega}{4 }\left[\tanh \frac{\Omega\tau}{2} (\x_f+\x_i)^2+ \coth \frac{\Omega\tau}{2}(\x_f-\x_i)^2\right]\right\},
\end{align}
with $\tform \equiv \sqrt{\omega/\hat q }$ and $\Omega \equiv (1+i)/(2\tform)$. In the absence of a medium, we get
\beq
\label{eq:K-vacuum}
 \mathcal{K}_0(\x_f,\x_i) \equiv  \left. \mathcal{K}(\x_f,\x_i) \right|_{\hat q \to 0}  = \frac{\omega}{2\pi i \tau} \, \exp \left[i \frac{\omega}{2 \tau }(\x_f-\x_i)^2\right]\,.
\eeq
The full expression reads then,
\begin{align} 
\label{eq:3-point-fct-ho-3}
&\tilde S^{(3)}(\u_f,\u_i,\v) \simeq \exp\left[-\frac{1}{4}\left(\frac{C_{F}}{N_c}-\frac{1}{4}\right) \hat q \, \v^2 \tau\right] \nn
& \times \frac{\omega \Omega}{2\pi i \, \sinh \Omega \tau} \, \exp\left\{ \frac{i\omega \Omega}{4 }\left[\tanh \frac{\Omega\tau}{2} (\x_f+\x_i)^2+ \coth \frac{\Omega\tau}{2}(\x_f-\x_i)^2\right]\right\} \,,
\end{align}
which can be approximated further if we assume that the formation time of the radiated gluon is the smallest time scale, i.e. $\tau < \tform  \ll (\hat q \v^2)^{-1}$. That is one can neglect the first factor in the above expression.
This leads to 
\begin{align} 
\label{eq:3-point-fct-ho-4}
\tilde S^{(3)}(\u_f,\u_i,\v) &\simeq \frac{\omega \Omega}{2\pi i \, \sinh \Omega \tau} \exp\left\{ \frac{i\omega \Omega}{4 }\left[\tanh \frac{\Omega\tau}{2} (\u_f+\u_i + \v)^2+ \coth \frac{\Omega\tau}{2}(\u_f-\u_i)^2\right]\right\} \,,
\end{align}
which in Fourier space  becomes
\begin{align}
\label{eq:3-point-fct-ho-FT2}
 \tilde S^{(3)}(\k,\q,\l) &= \int \rmd^2 \u_f  \rmd^2 \u_i  \rmd^2 \v \,\tilde S^{(3)}(\u_f,\u_i,\v)\, \rme^{-i \k\cdot \u_f+i \q\cdot \u_i - i \l\cdot \v} \nn &\simeq (2\pi)^2 \delta^{(2)}\left(\l - \frac{\k-\q}{2} \right) \nn
 &\times \frac{2\pi i }{ \omega \Omega \sinh \Omega \tau}\exp\left[-i \frac{(\k+\q)^2}{4 \omega\Omega \coth (\Omega\tau/2)} - i \frac{(\k-\q)^2}{4 \omega\Omega \tanh (\Omega\tau/2)}\right] \,.
\end{align}

%%%%%%%%%%%%%%%%%%%%%%%%%%%%%
\section{Suppressed real emission before interference}
\label{sec:before-interference}
%%%%%%%%%%%%%%%%%%%%%%%%%%%%%

Consider a real emission before an interference emission takes place, that is a radiative correction to $S_2$. In the large-$N_c$ limit, this leads to a correction to $S_2$ which contains the following color structure, 
\beq
\big\langle \rmtr \big(V_2^\dagger \tmat^a V_1 \big)\, \rmtr \big(V_2 \tmat^b V_1^\dagger \big)\, U_\r^{ab} \big\rangle \,,
\eeq
where all of the Wilson lines live in the interval $[\tf,\ti]$ and $U_\r^{ab}$ describes the gluon at a fluctuating transverse position $\r$. Whether this emission is a direct or an interference one only depends on the boundary conditions for $\r$.
However, rewriting the gluon propagator using the Fierz identity
\beq
U^{ab}_\r =\frac{1}{2} \rmtr \big(\tmat^a V_\r \tmat^b V_\r^\dagger \big) \,,
\eeq
we find that this color structure factorizes into a sextopole and a $1/N_c$-suppressed product of dipoles,
\beq
\big\langle \rmtr \big(V_2^\dagger \tmat^a V_1 \big)\, \rmtr \big(V_2 \tmat^b V_1^\dagger \big)\, U_\r^{ab} \big\rangle = \rmtr \big(V_\r^\dagger V_1 V_2^\dagger\, V_\r V_1^\dagger V_2 \big) - \frac{1}{N_c} \rmtr \big(V_1 V_2^\dagger \big) \, \rmtr \big( V_1^\dagger V_2 \big) \,,
\eeq
which are both sub-leading compared to virtual emissions (that do not connect the amplitude with the c.c.) in the large-$N_c$ limit.

\bibliographystyle{elsarticle-num}
\section*{\refname}
\bibliography{jetquenching}

\end{document}